\documentclass[aps,prd,preprint,nofootinbib]{revtex4}
\usepackage{bbm}
\usepackage{amsfonts}
\usepackage{mathrsfs}
\usepackage{graphicx}
\usepackage{amsmath}
\usepackage{amssymb}
\usepackage{bm}
\usepackage{bbm}
\usepackage{multirow}
\usepackage{epsfig}
\usepackage{graphicx}
\newcommand{\bqa}{\begin{eqnarray}}
\newcommand{\eqa}{\end{eqnarray}}
\newcommand{\beq}{\begin{equation}}
\newcommand{\eeq}{\end{equation}}
\newcommand{ \slashchar }[1]{\setbox0=\hbox{$#1$}   
   \dimen0=\wd0                                     
   \setbox1=\hbox{/} \dimen1=\wd1                   
   \ifdim\dimen0>\dimen1                            
      \rlap{\hbox to \dimen0{\hfil/\hfil}}          
      #1                                            
   \else                                            
      \rlap{\hbox to \dimen1{\hfil$#1$\hfil}}       
      /                                             
   \fi}                                             %
\begin{document}
\title{NLO QCD Corrections to ($\gamma^{*}\rightarrow Q \bar{Q}$)-Reggeon Vertex\\[7mm]}
\author{Li-Ping Sun$^{1}$, Long-Bin Chen$^{2}$, Cong-Feng
Qiao$^{3}$\footnote{E-mail: qiaocf@gucas.ac.cn } and Kuang-Ta
Chao$^{4}$\footnote{E-mail: ktchao@pku.edu.cn}\vspace{2mm}}

\affiliation{$^1$ School of Science, Beijing University of Civil
Engineering and Architecture, Beijing 100044,
China\\
$^2$ School of Physics and Materials Science, Guangzhou University,
Guangzhou 510006, China\\
$^3$ College of Physical Sciences, University of Chinese Academy of Sciences, Beijing 100049, China\\
$^4$ School of Physics and State Key Laboratory of Nuclear Physics and Technology, Peking University, and Center for High Energy Physics, Peking University, Beijing 100871, China}

\author{~\vspace{0.2cm}}


\begin{abstract}
{~\\[-3mm]} Next-to-leading order QCD corrections to the $\gamma^{*} \longrightarrow Q\bar{Q}-Reggeon$ vertex are
calculated, where $Q\bar{Q}$ denotes a heavy quark pair $c\bar c$ or $b\bar b$. The heavy quark mass effects on the
photon impact factor are found to be significant,  and hence may influence the results of high energy
photon-photon scattering and heavy quark pair leptoproduction.  In our NLO calculation, similar to the massless case, the ultraviolate(UV) divergences
are fully renormalized in the standard procedure, while the infrared (IR) divergences are regulated by  the parameter $\epsilon_{IR}$ in dimensional regularization. For the process $\gamma^* + q \to Q\bar{Q} + q$, we calculate all NLO coefficients in terms of $\epsilon_{IR}$, and find they are enhanced due to the heavy quark mass, as compared with the light quark case, and the enhancement factors increase rapidly as the quark mass increases. This might essentially indicate the quark mass effect, in spite of the absence of real corrections that are needed in a complete NLO calculation.
Moreover, unlike the $\gamma^{*}$ to massless-quark-Reggeon vertex,
the results in the present work may apply to the real photon case.

\vspace {7mm} \noindent
{PACS number(s): 12.38.Bx, 13.25.Gv, 14.40.Be }
\end{abstract}
\maketitle

\section{Introduction}

High energy diffractive scattering can be well described by the
Pomeron exchange mechanism \cite{pom1,pom2,pom3,pom4,pom5}. Within
the framework of quantum chromodynamics(QCD), the
Balitskii-Fadin-Kuraev-Lipatov(BFKL) Pomeron
\cite{bfkl1,bfkl2,bfkl3}, the so-called hard Pomeron, is viewed as
the Green's function of two interacting Reggeized gluons with vacuum
quantum number, which is evaluated by resuming the leading energy
logarithms in perturbative QCD(pQCD) applicable regime, e.g., all
$(\alpha_s \text{ln} s)^n $ terms in leading logarithmic
approximation(LLA) and all $\alpha_s (\alpha_s \text{ln} s)^n $
terms in the next-to-leading approximation(NLA). People noticed that
the virtual photon-photon scattering is an ideal process
\cite{virpho1,virpho2} to testify the BFKL predictions and hence the
pQCD calculation reliability, which is highly expected in the study
of strong interaction physics. The cross section
$\sigma_{tot}^{\gamma^{*} \gamma^{*}}$ is calculable in the BFKL
approach with relatively high precision and can be realized in
experiment through a measurement of the reaction $e^{+}
e^{-}\rightarrow e^{+} e^{-} + X$ by tagging the outgoing leptons at
high energy electron-positron colliders, like LEP and more
optimistically CEPC \cite{cepc} or ILC \cite{ilc} in the future.
Moreover, the recently proposed electron-ion colliders like EIC
\cite{eic} and EicC \cite{eicc} may also be good places to study the
single diffractive process and hence on small-x physics at lower
energies.

Till now, the leading order(LO) BFKL predictions for $e^{+}
e^{-}\rightarrow e^{+} e^{-} + X$ process have been confronted to
the LEP data \cite{lep1,lep2,lep3}, in which the experimental
measurement is found above the two-gluon-exchange model result,
while below the leading order BFKL Pomeron prediction. Now that the
next-to-leading order(NLO) corrections to BFKL kernel is ready and
numerically big and negative \cite{bfkl3,nlb2}, the higher order
corrections may lower the the BFKL Pomeron exchange calculation and
approach to the experimental measurement. However, in the NLO BFKL
calculations, there is still a task remaining, calculating the NLO
corrections to the coupling of the BFKL Pomeron and the external
photons, which is called photon impact factor\cite{im1,im2,im3}.

The photon impact factor can be obtained by calculating the
discontinuity of the process $\gamma^{*}+~reggon \longrightarrow
\gamma^{*}+~reggon$ process in light of the optical theorem, which
tells that the discontinuity is proportional to the scattering
amplitude, $\gamma^{*}+~reggeon\longrightarrow Q\bar{Q}$, squared.
The reggon, is identified with the elementary t-channel gluons. The
NLO corrections to the photon impact factor with massless internal
quarks were given in Ref.\cite{qiao}, while here the $Q(\bar{Q})$
are massive quarks, charm or bottom quarks. In the NLO calculation,
in analogous to Ref.\cite{qiao}, we calculate the NLO corrections to
$\gamma^{*}+~reggeon\longrightarrow Q\bar{Q}$ amplitude on for
example left-hand side of the discontinuity line, and the leading
$Q\bar{Q}g$ intermediate state on both sides of the discontinuity
line. The calculation of NLO QCD correction hence proceeds in three
steps: (i) calculate the NLO corrections to the
$\gamma^{*}+~reggeon\longrightarrow Q\bar{Q}$ vertex. (ii) calculate
the leading order vertex $\gamma^{*}+~reggeon \longrightarrow
Q\bar{Q}g$, and (iii) carry out the integration over the phase space
of the intermediate states. In fact, here the calculation procedure is similar to the massless case performed in Ref.\cite{qiao}. In this paper, as in \cite{qiao}, we give the results of
the first step, in which the vertex is extracted from the scattering
process $\gamma^{*}+~q\longrightarrow Q\bar{Q}+q$ in the high energy
limit.

\section{Technical preliminaries}\label{II}

\begin{figure}[!hbtp]
\centering
\includegraphics[scale=0.8]{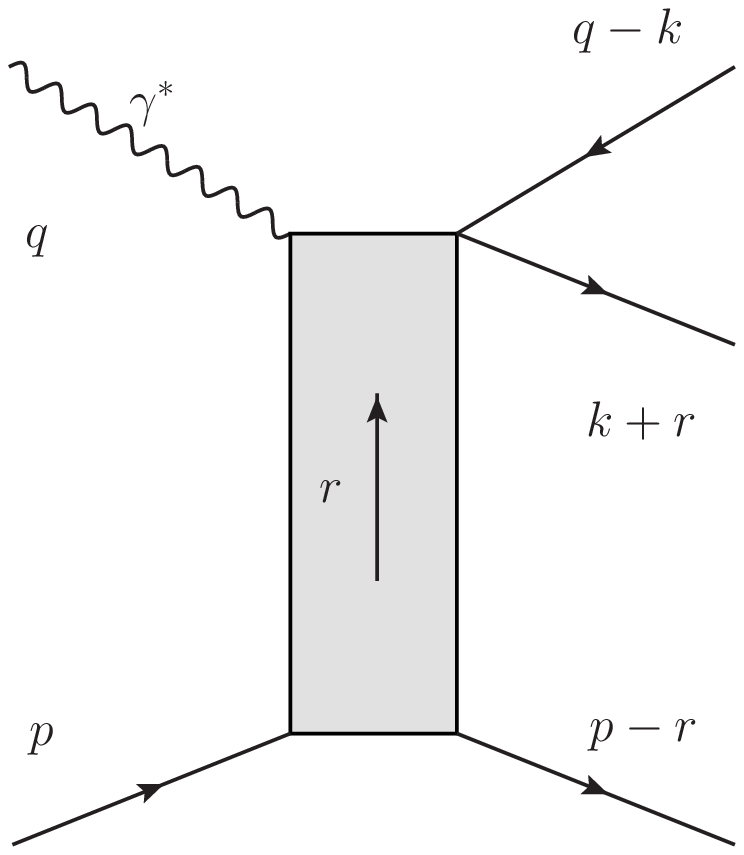}%
\caption{\small Kinematics of the process $\gamma^* + q \to Q\bar{Q}
+ q$. } \label{kinematics}
\end{figure}

The kinematics of the $\gamma^* + q \to Q\bar{Q} + q$ process is
illustrated Fig.\ref{kinematics}, where $q$ and $p$ are the
4-momenta of photon and incident quark respectively. In the
calculation, $\varepsilon$ represents for the polarization
vector of the photon, $s$ for the collision energy of the $\gamma^*
q$ system, and $m$ for the mass of the heavy quark. The Lorentz
invariants appeared include $s=(q+p)^2$, $Q^2=-q^2$, $t_{a}=k^2$,
$t_{b}=(q-k-r)^2$, $M^2=(q+r)^2$, $t=r^2$, and $x= Q^2/2p\cdot q$,
the Bjorken scaling variable.

The momenta $k$ and $r$ can be decomposed in the Sudakov form, i.e,

\begin{eqnarray}
k &=&\alpha q^\prime+\beta p+k_{\bot}\ , \\
r &=& \frac{t}{s} q^\prime -\frac{t_{a}+t_{b}-2m^2}{s} p + r_{\bot}\
,
\end{eqnarray}
with $q^\prime =q+xp$, ${q^{\prime}}^2=p^2=0$ and $\beta s
=\frac{k_{\bot}^2+m^2}{1-\alpha}-Q^2$.

\begin{figure}[tb]
  \begin{center}
    \epsfig{file=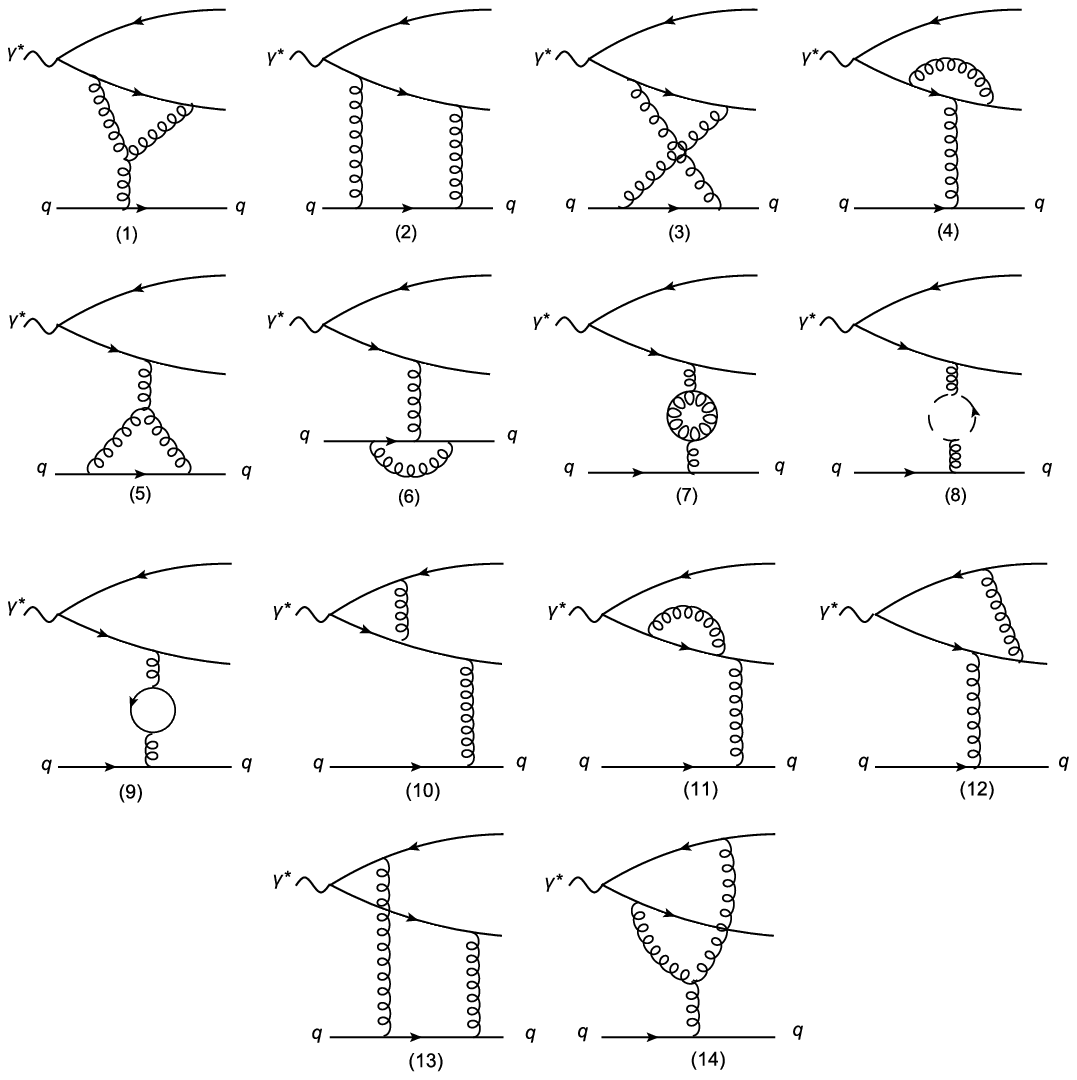,width=6in}
  \end{center}
  \caption{Feynman diagrams for the process $\gamma^* + q \to Q\bar{Q} +
    q$.}
  \label{graphs}
\end{figure}

The typical Feynman diagrams for NLO corrections are shown in
Fig.\ref{graphs}. Of these diagrams, except Fig.\ref{graphs}.14, the
upper part quark-antiquark exchange diagrams are implied. For the
color octet t-channel configuration, the sum of all diagrams has to
be antisymmetric if we interchange quark and antiquark: $k
\rightarrow q-k-r$, $\lambda\rightarrow\lambda^{\prime}$, where
$\lambda$, $\lambda^{\prime}$ are the helicities of the quark and
antiquark respectively. In particular, the "box" graph shown in Fig. 3.14
has to be antisymmetric by itself. Throughout the calculation,
Feynman gauge is employed, and for the t-channel gluons the metric
tensor is decomposed into \bqa
g_{\mu\nu}=\frac{2}{s}(p_{\mu}q^{\prime}_{\nu}+p_{\nu}q^{\prime}_{\mu})
+g^{\bot}_{\mu\nu}\eqa as usual. Note, in practical calculations, the
transverse term is not taken into account, whose contributions are
suppressed by powers of $s$. We use the helicity formalism as well,
and then our results can be expressed in
terms of the following matrix elements similar to \cite{qiao}, i.e.,
\bqa
\label{HTdef} H_{T}^a&=& \bar{u}(k+r, \lambda) \not\!{p} \not\!{k}
\not\!{\varepsilon}\;
\lambda^av(q-k,\lambda') \; , \\
\label{HEdef} H_{\varepsilon}^a&=& \bar{u}(k+r, \lambda)
\not\!{\varepsilon}\;  \lambda^a
v(q-k,\lambda') \; , \\
\label{HKdef} H_{k}^a&=& \bar{u}(k+r, \lambda) \not\!{k}\; \lambda^a
v(q-k,\lambda')
\; , \\
\label{HPdef} H_{p}^a&=& \bar{u}(k+r, \lambda) \not\!{p}\; \lambda^a
v(q-k,\lambda')
\; , \\
\label{HKEdef} H_{k\varepsilon}^a&=& \bar{u}(k+r, \lambda)
\not\!{k}\not\!{\varepsilon}\; \lambda^a v(q-k,\lambda')
\; , \\
\label{HPEdef} H_{p\varepsilon}^a&=& \bar{u}(k+r, \lambda)
\not\!{p}\not\!{\varepsilon}\; \lambda^a v(q-k,\lambda')
\; , \\
\label{HPKdef} H_{pk}^a&=& \bar{u}(k+r, \lambda)
\not\!{p}\not\!{k}\; \lambda^a v(q-k,\lambda')
\; , \\
\label{HMdef} H_{m}^a&=& \bar{u}(k+r, \lambda) \; \lambda^a
v(q-k,\lambda'). \eqa Here $\lambda^a$ is the generators of the
color group. Note that the last four helicity matrix elements do not exist in \cite{qiao},
which always come up with the factor of $m$.

Take high energy limit in the calculation, where
\bqa t,
Q^2, t_a, t_b, M^2, m^2 \ll s,
\eqa
and do not impose any restrictions on the remaining invariants, we then obtain the
following amplitude
\bqa \label{eq:16} T = g^2\;T^{(0)} + g^4\;T^{(1)},
\eqa
with
\bqa T^{(0)} = \Gamma_{\gamma^* \to q\bar{q}}^{(0),a}\; \frac{2s}{t}
\; \Gamma_{qq}^{(0),a},
\eqa and \bqa \Gamma_{qq}^{(0),a}&=&
\frac{1}{s} \bar{u}(p-r, \,\lambda^{\prime}) \not\!{q}' \lambda^a
u(p,\,\lambda)\ . \eqa
The only unknown piece of the NLO amplitude is
$T^{(1)}$, which corresponds to the processes in Fig.\ref{graphs}.1 - \ref{graphs}.14
and will be calculated analytically in the following.

\section{Calculation methods}

The method we use in our calculation will be briefly described
in the following before presenting the explicit results.

In our calculation the MATHEMATICA package FeynArts \cite{feynarts}
is applied to generate the Feynman diagrams and amplitudes that are
relevant to our process. We use FeynCalc \cite{feyncalc} to
calculate and simplify the amplitudes. Color in the t-channel is
projected onto the antisymmetric octet as done in \cite{qiao}. After
simplifying the Dirac matrix and employing the Dirac equation of motion, we
express the amplitudes as the combination of helicity matrix
elements and Passarino-Veltman integrals. For integrals that contain
divergences, we separate the divergence part from the finite one.
The high energy limit is taken through out our calculation, i.e. $t,
Q^2, t_a, t_b, M^2, m^2 \ll s$.

In this work an extra scale $m$ exists in comparison to \cite{qiao}, and
it is too lengthy and redundancy to turn all the Passarino-Veltman
integrals into scalar integrals or logarithms and dilogarithms
functions. For amplitudes that do not contain divergences and
high energy scale $s$, we express them as the combination of
Passarino-Veltman integrals and helicity matrix elements. The
numerical values of these integrals will be evaluated
by Looptools \cite{looptools}. For the pentagon diagram
Fig.\ref{graphs}.13, the reduction method given in
\cite{denner} is employed to reduce the corresponding integrals to box integrals.
In the end we find our result agrees with that in \cite{qiao} when taking the $m\rightarrow 0$ limit.

\section{Analytic Results}

The NLO amplitudes in $T^{(1)}$ of our concern can be expressed in terms of
different diagrams:
\bqa
T^{(1)}
&=&\sum_{i=1}^{13}(A_{i}+\bar{A}_{i})+A_{14}\ .
\eqa
Here, the subscripts $i$ denote diagrams in Fig.\ref{graphs}, and the amplitudes
$\bar{A}_{i}$ represent those with quark-antiquark interchange in ${A}_{i}$.

\subsection{Results of two- and three-point diagrams}

We categorize the diagrams similar to \cite{qiao}. The results of
Figs.\ref{graphs}.1 and its conjugate one are given below, which
are split into divergent and finite parts in $\mathrm{\overline{MS}}$ scheme:
\begin{align}
A_{1}\,&=\, \frac{N_c}{2}\frac{e
e_f}{(4\pi)^{2-\epsilon}}\frac{2}{t(t_a-m^2)}\Gamma_{qq}^{(0),a}
\bigg\{\frac{3(H_{T}^a+m H_{p\varepsilon}^a)c_\Gamma}{\epsilon}
+H_{T}^a
\big[6m^2C_0(1)+12C_{00}(1)_{fin}\nonumber\\
&+(5m^2-2t+3t_a)C_1(1)+(3m^2-t+t_a)C_2(1)+2t_aC_{11}(1)
+2(m^2-t+t_a)C_{12}(1) \nonumber\\
&+2m^2C_{22}(1)-2\big]+\alpha s
H_{\varepsilon}^a\big[2m^2C_1(1)+(3m^2-t_a)C_2(1)+2t_a
C_{11}(1)+2m^2C_{22}(1)\nonumber\\
&+2(t_a+m^2)C_{12}(1) \big]+m
H_{p\varepsilon}^a\big[3(t_a+m^2)C_0(1)+12C_{00}(1)_{fin}+2(m^2-t+3t_a)C_1(1)\nonumber\\
&+ (3m^2-2t+5t_a)C_2(1)+2t_aC_{11}(1)
+2(m^2-t+t_a)C_{12}(1)+2m^2C_{22}(1)-2 \big]\nonumber\\
&+2\alpha s
mH_{k\varepsilon}^a\big[C_1(1)+C_2(1)+C_{11}(1)+2C_{12}(1)+C_{22}(1)
\big]\bigg\}\ .
\end{align}
Here $C_i(1)=C_i(t_a,t,m^2,m^2,0,0)$ with $i$ the coefficient index.
The definitions of all the loop integrals such as $C_0$,
$C_{ij}$,$D_0$, $D_{i,j,k}$ are standard and given in the Appendix
for reference. The subindex $fin$ means the finite part of the loop
integrals. Note, the above result does not contain any infrared
divergences, while the ultraviolet divergence remains only in
$C_{00}(1)$, which can be properly regulated. Numerical evaluation
of the finite part may be performed by means of LoopTools
\cite{looptools} in $\mathrm{\overline{MS}}$ scheme. The $C_i$
functions can be reduced into a certain scalar one-loop integrals
and then expressed as the combination of logarithm and dilogarithm
functions. Nevertheless, such kind of reduction may yield a lengthy
result and hard to read. Due to this reason, we do not perform
further reduction on $C_i$ functions, its numerical evaluations are
carried by LoopTools directly.

The result of the conjugate diagram of Figs.\ref{graphs}.1 can be expressed as
\begin{align}
\bar{A}_{1}\,&=\,\frac{N_c}{2}\frac{e
e_f}{(4\pi)^{2-\epsilon}}\frac{2}{t(t_b-m^2)}
\Gamma_{qq}^{(0),a}\bigg\{\frac{3(H_{T}^a+m
H_{p\varepsilon}^a-s H_{\varepsilon}^a +2(H_{k}^a-m
H_{m}^a)\varepsilon\cdot p+2H_{p}^a\varepsilon\cdot
r)c_\Gamma}{\epsilon}\nonumber\\
&+(H_{T}^a+2H_{p}^a\varepsilon\cdot r + 2H_{k}^a\varepsilon\cdot
p)\big[6m^2C_{0}(\bar{1})+12C_{00}(\bar{1})_{fin}+2(3m^2-t+t_b)C_{1}
(\bar{1})\nonumber\\
&+(5m^2-2t+3t_b^2)C_{2}(\bar{1})+2m^2C_{11}(\bar{1})
+2(m^2-t+t_b)C_{12}(\bar{1})+2t_bC_{22}(\bar{1})-2\big]\nonumber\\
&+sH_{\varepsilon}^a\big[-6m^2C_{0}(\bar{1})+(3(\alpha-3)m^2
+2t-(1+\alpha)t_b)C_{1}(\bar{1})-12C_{00}(\bar{1})_{fin}\nonumber\\
&+((2\alpha-7)m^2+2t-3t_b)
C_{2}(\bar{1})+2(\alpha-2)m^2C_{11}(\bar{1})
+2(\alpha-2)t_bC_{22}(\bar{1})\nonumber\\
&+2((\alpha-2)m^2+t+(\alpha-2)t_b)C_{12}(\bar{1})+2\big]\nonumber\\
&+2(\alpha-1)s m (H_{k\varepsilon}^a+H_{m}^a \varepsilon\cdot r)
\big[C_{1}(\bar{1})+C_{2}(\bar{1})+C_{11}(\bar{1})
+2C_{12}(\bar{1})+C_{22}(\bar{1})\big]\nonumber\\
&+m( H_{p\varepsilon}^a-2\varepsilon\cdot p
H_{m}^a)\big[(9m^2-3t_b)C_{0}(\bar{1})+12C_{00}(\bar{1})_{fin}+(9m^2-2t-t_b)C_{1}
(\bar{1})\nonumber\\
&+2(4m^2-t)C_{2}(\bar{1})+2m^2C_{11}(\bar{1})
+2(m^2-t+t_b)C_{12}(\bar{1})+2t_bC_{22}(\bar{1}) -2\big] \bigg\}\ .
\end{align}
Here, $C_i(\bar{1})=C_i(m^2,t,t_b,m^2,0,0)$.

The calculation procedure of Fig.\ref{graphs}.4 is similar to that
of Fig.1, and its result reads:
\begin{align}
A_{4}\,&=\,-\frac{1}{2N_c}\frac{e
e_f}{(4\pi)^{2-\epsilon}}\frac{4}{t(t_a-m^2)}\Gamma_{qq}^{(0),a}
\bigg\{\frac{(H_{T}^a+mH_{p\varepsilon}^a)c_\Gamma}{2\epsilon}+
H_{T}^a\big[(m^2-t+t_a)C_{0}(4)\nonumber\\
&+(3m^2-t+t_a)C_{1}(4)+(2m^2-t+2t_a)C_{2}(4)+2C_{00}(4)_{fin}+m^2C_{11}(4)\nonumber\\
&+(m^2-t+t_a)C_{12}(4) +t_aC_{22}(4)-1\big]+ \alpha s
H_{\varepsilon}^a\big[(m^2-t_a)C_{0}(4)-t_aC_{1}(4)\nonumber\\
& +(m^2-2t_a)2C_{2}(4)-m^2C_{11}(4)-(t_a+m^2)C_{12}(4)-t_aC_{22}(4)
\big]\nonumber\\
&+m H_{p\varepsilon}^a\big[(2t_a-t)C_{0}(4)+(2m^2-t+2t_a)C_{1}(4)
+(m^2-t+3t_a)C_{2}(4)\nonumber\\
&+2C_{00}(4)_{fin}+m^2C_{11}(4)+(m^2-t+t_a)C_{12}(4)+t_aC_{22}(4)-1\big]\nonumber\\
&-\alpha s m
H_{k\varepsilon}^a\big[C_{1}(4)+C_{2}(4)+C_{11}(4)+2C_{12}(4)+C_{22}(4)
\big]\bigg\}\ .
\end{align}
where $C_i(4)=C_i(m^2,t,t_a,0,m^2,m^2)$.

The result of the conjugate diagram of Fig.\ref{graphs}.4 is:
\begin{align}
\bar{A}_{4}\,&=\,-\frac{1}{2N_c}\frac{e
e_f}{(4\pi)^{2-\epsilon}}\frac{4}{t(t_b-m^2)}\Gamma_{qq}^{(0),a}
\bigg\{\nonumber\\
&\frac{(H_{T}^a+mH_{p\varepsilon}^a-s H_{\varepsilon}^a
+2(H_{k}^a-mH_{m}^a)\varepsilon\cdot p+2H_{p}^a\varepsilon\cdot
r)c_\Gamma}{2\epsilon}\nonumber\\
&+H_{T}^a\big[(m^2-t+t_b)C_{0}(\bar{4})+(3m^2-t+t_b)C_{1}(\bar{4})
+(2m^2-t+2t_b)C_{2}(\bar{4})\nonumber\\
&+2C_{00}(\bar{4})_{fin}+m^2C_{11}(\bar{4})+(m^2-t+t_b)C_{12}(\bar{4})
+t_bC_{22}(\bar{4})-1\big]\nonumber\\
&+sH_{\varepsilon}^a\big[((\alpha-2)m^2+t-\alpha t_b)C_{0}(\bar{4})
+(t-3m^2-\alpha t_b)C_{1}(\bar{4})\nonumber\\
&+((\alpha-3)m^2+t-2\alpha
t_b)C_{2}(\bar{4})-2C_{00}(\bar{4})_{fin}-\alpha m^2C_{11}(\bar{4})\nonumber\\
& +(t-\alpha m^2-\alpha t_b)C_{12}(\bar{4})-\alpha
t_bC_{22}(\bar{4})+1\big]\nonumber\\
&+2(H_{k}^a \varepsilon\cdot p+H_{p}^a\varepsilon\cdot
r)\big[(m^2-t+t_b)C_{0}(\bar{4})+2C_{00}(\bar{4})_{fin}\nonumber\\
&+(3m^2-t+t_b)C_{1}(\bar{4})+(2m^2-t+2t_b)C_{2}(\bar{4})
+m^2C_{11}(\bar{4})\nonumber\\
&+(m^2-t+t_b)C_{12}(\bar{4})+t_bC_{22}(\bar{4})-1
\big]\nonumber\\
&+m(H_{p\varepsilon}^a+2H_{m}^a\varepsilon\cdot
p)\big[(2m^2-t)C_{0}(\bar{4})+(4m^2-t)C_{1}(\bar{4})+(3m^2-t+t_b)C_{2}(\bar{4})\nonumber\\
&+2C_{00}(\bar{4})_{fin}+m^2C_{11}(\bar{4})+(m^2-t+t_b)C_{12}(\bar{4})
+t_bC_{22}(\bar{4})-1\big]\nonumber\\
&+(1-\alpha)m(H_{k\varepsilon}^a+2H_{m}^a\varepsilon\cdot
r)\big[C_{1}(\bar{4})+C_{2}(\bar{4})+C_{11}(\bar{4})+2C_{12}(\bar{4})+C_{22}(\bar{4})
\big] \bigg\}.
\end{align}
with $C_i(\bar{4})=C_i(m^2,t,t_b,0,m^2,m^2)$.

The Fig.\ref{graphs}.5 and Fig.\ref{graphs}.6 diagrams
belong to the NLO corrections to the lower vertex
$\Gamma_{qq}^{(1),a}$. Their results are
\begin{align}
A_{5}\,&=\,-N_c\frac{e
e_f}{(4\pi)^{2-\epsilon}}\frac{2}{t(t_a-m^2)}\Gamma_{qq}^{(0),a}
(H_{T}^a+mH_{p\varepsilon}^a)c_\Gamma\bigg\{\frac{(-t)^{-\epsilon}}{2\epsilon}+1\bigg\},
\end{align}
and
\begin{align}
A_{6}\,&=\,\frac{1}{2N_c}\frac{e
e_f}{(4\pi)^{2-\epsilon}}\frac{2}{t(t_a-m^2)}\Gamma_{qq}^{(0),a}
(H_{T}^a+mH_{p\varepsilon}^a)c_\Gamma(-t)^{-\epsilon}
\bigg\{\frac{2}{\epsilon^2}+\frac{3}{\epsilon}+8\bigg\}\ .
\end{align}
In the zero mass limit, it is obvious that these two amplitudes
are just the eqs. $(35)$ and $(36)$ in \cite{qiao}.

Figs.\ref{graphs}.7 - \ref{graphs}.9 contribute to both
the upper and lower vertices, and we find
\begin{align}
A_{7+8}\,&=\,\frac{e
e_f}{(4\pi)^{2-\epsilon}}\frac{2N_cc_\Gamma(-t)^{-\epsilon}}{t(t_a-m^2)}
\Gamma_{qq}^{(0),a}(H_{T}^a+mH_{p\varepsilon}^a)\bigg(\frac{5}{3\epsilon}+\frac{31}{9}
\bigg),
\end{align}
and
\begin{align}
A_{9}\,&=\,-n_f\frac{e
e_f}{(4\pi)^{2-\epsilon}}\frac{2c_\Gamma(-t)^{-\epsilon}}{t(t_a-m^2)}
\Gamma_{qq}^{(0),a}(H_{T}^a+mH_{p\varepsilon}^a)\bigg(\frac{2}{3\epsilon}
+\frac{10}{9}\bigg)\ .
\end{align}

The results of conjugate diagrams of Fig.\ref{graphs}.5 -
\ref{graphs}.9 can readily be obtained by substituting
$(H_{T}^a+mH_{p\varepsilon}^a)$ with
$(H_{T}^a+mH_{p\varepsilon}^a-sH_{\varepsilon}^a+2(H_{k}^a-mH_{m}^a)\varepsilon\cdot
p+2H_{p}^a\varepsilon\cdot r)$  and making the replacement
$t_a\rightarrow t_b$.

The result of vertex correction diagram Fig.\ref{graphs}.10 can be
expressed as follows:
\begin{align}
A_{10}\,&=\,-\frac{C_F e
e_f}{(4\pi)^{2-\epsilon}}\frac{4}{t(t_a-m^2)}\Gamma_{qq}^{(0),a}
\bigg\{\frac{-(H_{T}^a+mH_{p\varepsilon}^a)c_\Gamma}{2\epsilon}+
H_{T}^a\big[m^2C_{0}(10)+Q^2C_{2}(10)\nonumber\\
&+(m^2+Q^2+t_a)C_{1}(10)-2C_{00}(10)_{fin}-m^2C_{11}(10)
+(Q^2+t_a-m^2)C_{12}(10)\nonumber\\
&+Q^2C_{22}(10)+1\big]+mH_{p\varepsilon}^a\big[m^2C_{0}(10)
+(Q^2+2t_a)C_{1}(10)+Q^2C_{2}(10)\nonumber\\
&-2C_{00}(10)_{fin}-m^2C_{11}(10)+(Q^2-m^2+t_a)C_{12}(10)+Q^2C_{22}(10)+1\big]\nonumber\\
&+2\varepsilon\cdot
k\big[(m(2mH_{p}^a+H_{pk}^a)-t_aH_{p}^a)C_{1}(10)
+m(mH_{p}^a+H_{pk}^a)C_{11}(10)\nonumber\\
&+H_{p}^a(m^2-t_a)C_{12}(10) \big]\bigg\}.
\end{align}
Here $C_i(10)=C_i(m^2,t_a,-Q^2,m^2,0,m^2)$.

The NLO amplitude of the conjugate diagram of Fig.\ref{graphs}.10 is
\begin{align}
\bar{A}_{10}\,&=\,-\frac{C_F e
e_f}{(4\pi)^{2-\epsilon}}\frac{4}{t(t_b-m^2)}\Gamma_{qq}^{(0),a}
\bigg\{\nonumber\\
&-\frac{(H_{T}^a+mH_{p\varepsilon}^a-sH_{\varepsilon}^a+2(H_{k}^a-mH_{m}^a)\varepsilon\cdot
p+2H_{p}^a\varepsilon\cdot r)c_\Gamma}{2\epsilon}\nonumber\\
& +(H_{T}^a-sH_{\varepsilon}^a+2H_{k}^a\varepsilon\cdot p)\big[
m^2C_{0}(\bar{10})+(m^2+Q^2+t_b)C_{1}(\bar{10})+Q^2C_{2}(\bar{10})\nonumber\\
&-2C_{00}(\bar{10})_{fin}-m^2C_{11}(\bar{10})+(Q^2-m^2+t_b)C_{12}(\bar{10})
+Q^2C_{22}(\bar{10})+1\big]\nonumber\\
&+mH_{p\varepsilon}^a\big[m^2C_{0}(\bar{10})+(2m^2+Q^2)C_{1}(\bar{10})
+Q^2C_{2}(\bar{10})-2C_{00}(\bar{10})_{fin}-m^2C_{11}(\bar{10})\nonumber\\
&+(Q^2-m^2+t_b)C_{12}(\bar{10})
+Q^2C_{22}(\bar{10})+1\big]+2H_{pk}^a\big[C_{1}(\bar{10})
+C_{11}(\bar{10})\big]\times\nonumber\\
&(\varepsilon\cdot r+\varepsilon\cdot
k)+2mH_{p}^a\big[((2m^2-t_b)C_{1}(\bar{10})
+m^2C_{11}(\bar{10})+(m^2-t_b)C_{12}(\bar{10}))\varepsilon\cdot k\nonumber\\
&+(m^2C_{0}(\bar{10})+(3m^2+Q^2)C_{1}(\bar{10})+Q^2C_{2}(\bar{10})
-2C_{00}(\bar{10})_{fin}+Q^2C_{12}(\bar{10})\nonumber\\
&+Q^2C_{22}(\bar{10})+1)\varepsilon\cdot
r\big]+2mH_{m}^a\big[(-m^2C_{0}(\bar{10})-(2m^2+Q^2)C_{1}(\bar{10})
-Q^2C_{2}(\bar{10})\nonumber\\
&+2C_{00}(\bar{10})_{fin}+m^2C_{11}(\bar{10})+(m^2-Q^2-t_b)C_{12}(\bar{10})
-Q^2C_{22}(\bar{10})-1)\varepsilon\cdot p\nonumber\\
&-s(C_{1}(\bar{10})+C_{11}(\bar{10}))(\varepsilon\cdot
r+\varepsilon\cdot k)\big] \bigg\}.
\end{align}
with $C_i(\bar{10})=C_i(m^2,t_b,-Q^2,m^2,0,m^2)$.

The result of the quark self-energy diagram Fig.\ref{graphs}.11 is
\begin{align}
A_{11}\,&=\,\frac{e
e_f}{(4\pi)^{2-\epsilon}}\frac{2C_Fc_\Gamma}{t(t_a-m^2)^2}
\Gamma_{qq}^{(0),a}\bigg\{\frac{(7m^2-t_a)H_T^a+2(t_a
+2m^2)mH_{p\varepsilon}^a}{\epsilon}+\ln(m^2)[\nonumber\\
&(t_a-7m^2)H_T^a-2(t_a+2m^2)mH_{p\varepsilon}^a]+\frac{4t_a(t_a+m^2)mH_{p\varepsilon}^a
-(m^4-10t_am^2+t_a^2)H_T^a}{t_a}\nonumber\\
&+\frac{\ln(1-\frac{t_a}{m^2})(t_a-m^2)((m^4-6t_am^2+t_a^2)H_T^a
-2t_a(t_a+m^2)mH_{p\varepsilon}^a)}{t_a^2}\bigg\}.
\end{align}
and the amplitude of the conjugate diagram of Fig.3.11 reads
\begin{align}
\bar{A}_{11}\,&=\,\frac{e
e_f}{(4\pi)^{2-\epsilon}}\frac{2C_Fc_\Gamma}{t(t_b-m^2)^2}
\Gamma_{qq}^{(0),a}\bigg\{\frac{(7m^2-t_b)(H_T^a-sH_{\epsilon}^a+2H_k^a\varepsilon\cdot
p+2H_p^a\varepsilon\cdot r)}{\epsilon}\nonumber\\
&+\frac{2(5m^2-2t_b)m(H_{p\varepsilon}^a-2H_m^a\varepsilon\cdot
p)}{\epsilon}+\frac{1}{t_a}\big[(H_T^a-sH_{\epsilon}^a+2H_k^a\varepsilon\cdot
p+2H_p^a\varepsilon\cdot r\nonumber\\
&)(-t_b(7m^2-t_b)\ln(m^2)-\frac{m^2-t_b}{t_b}(m^4-6m^2t_b
+t_b^2)\ln(\frac{m^2-t_b}{m^2})\nonumber\\
&-m^4+10m^2t_b-t_b^2)+2m(H_{p\varepsilon}^a-2H_m^a\varepsilon\cdot
p)(-t_b(5m^2-2t_b)\ln(m^2)\nonumber\\
&-\frac{m^2-t_b}{t_b}(m^4-5m^2t_b+2t_b^2)\ln(\frac{m^2-t_b}{m^2})-m^4+8m^2t_b-3t_b^2)\big]\bigg\}\ .
\end{align}

\subsection{The Box diagrams}

Here, we present the results of box diagrams. In the calculation of diagrams Fig.\ref{graphs}.2 and
Fig.\ref{graphs}.3, the four-point integrals have to be concerned. After taking the high energy limit and eliminating the $s$ suppressed
terms, in the end the results turn out to be quit simple, that is
\begin{align}
A_{2+3}\,&=\,\frac{e
e_fN_c}{(4\pi)^{2-\epsilon}}\frac{c_\Gamma}{t(t_a-m^2)}
\Gamma_{qq}^{(0),a}(H_{T}^a+mH_{p\varepsilon}^a)\big\{\frac{(-t)^{-\epsilon}}{\epsilon^2}
-\frac{3(m^2)^{-\epsilon}}{2\epsilon^2}+\frac{(m^2)^{-\epsilon}}{\epsilon}
\big(\nonumber\\
&\ln(-\frac{\alpha s}{m^2})+\ln(\frac{\alpha
s}{m^2})+\ln(\frac{-t}{m^2})-\ln(1-\frac{t_a}{m^2})\big)
-\ln(-\frac{t}{m^2})\big[\ln(-\frac{\alpha s}{m^2})+\ln(\frac{\alpha
s}{m^2})\big]\nonumber\\
&-\ln(1-\frac{t_a}{m^2})^2-\frac{3\pi^2}{4}+C_{0}(1) \big\}.
\end{align}
Here, $C_0(1)=C_0(t_a,t,m^2,m^2,0,0)$, as in the $A_1$.

The results of diagrams conjugating to Fig.\ref{graphs}.2 and
Fig.\ref{graphs}.3 can be obtained by taking the following
replacements:
\\$t_a\rightarrow t_b$, $\alpha\rightarrow 1-\alpha$,
$(H_{T}^a+mH_{p\varepsilon}^a)\rightarrow
(H_{T}^a+mH_{p\varepsilon}^a-sH_{\varepsilon}^a+2(H_{k}^a-mH_{m}^a)\varepsilon\cdot
p+2H_{p}^a\varepsilon\cdot r)$.

Note that there are $\text{ln} s$ terms in the amplitudes of
Figs.2.2 and 2.3, and also their conjugate partners.

Next, we give the result of adjacent box as shown in Fig.\ref{graphs}.12.
\begin{align}
A_{12}\,&=\,\Gamma_{qq}^{(0),a}\frac{2s}{t}\bigg(\frac{e
e_f}{s}\bigg)\bigg(-\frac{1}{2N_c}\bigg)\frac{2}{(4\pi)^{2-\epsilon}}\bigg\{
\frac{c_\Gamma}{\epsilon}A_{12}^{-1}+A_{12}^0\bigg\}\ .
\end{align}
in which the divergent part $A_{12}^{-1}$ is
\begin{align}
A_{12}^{-1}\,&=\,-(H_{T}^a+mH_{p\varepsilon}^a)\frac{M^2x_r\ln(x_r)}{m^2(1-x_r^2)},
\end{align}
with \bqa
x_r=-\frac{1-\sqrt{1-\frac{4m^2}{2m^2+M^2}}}{1+\sqrt{1-\frac{4m^2}{2m^2+M^2}}}\
, \eqa and $M^2=(q+r)^2$, defined in the above paragraph.

The finite term $A_{12}^{0}$ is given in the Appendix. The amplitude
of the conjugate diagram of Fig.\ref{graphs}.12 goes as
\begin{align}
\bar{A}_{12}\,&=\,\Gamma_{qq}^{(0),a}\frac{2s}{t}\bigg(\frac{e
e_f}{s}\bigg)\bigg(-\frac{1}{2N_c}\bigg)\frac{2}{(4\pi)^{2-\epsilon}}\bigg\{
\frac{c_\Gamma}{\epsilon}\bar{A}_{12}^{-1}+\bar{A}_{12}^0\bigg\}\ ,
\end{align}
where the divergent term
\begin{align}
\bar{A}_{12}^{-1}=-\frac{M^2x_r\ln(x_r)}{m^2(1-x_r^2)}
\big[H_{T}^a+mH_{p\varepsilon}^a-sH_{\varepsilon}^a
+2(H_{k}^a-mH_{m}^a)\varepsilon\cdot p+2H_{p}^a\varepsilon\cdot
r\big]\ .
\end{align}
The finite piece is also given in the Appendix.

The opposite box diagram Fig.\ref{graphs}.14 does not contain any
divergence, its lengthy analytic expression is presented in the Appendix.

\subsection{The Pentagon diagram}
In this subsection we deal with the pentagon diagram
Fig.\ref{graphs}.13. The calculation procedure is complicated, and is performed by computer algebra. Due to the fact
that the integrals depend upon the large scale $s$,  they can be greatly simplified in the high energy
limit.  The results are as follows:
\begin{align}
A_{13}\,&=\,\Gamma_{qq}^{(0),a}\frac{N_c}{2}\frac{ee_f}{t(4\pi)^{2-\epsilon}}\bigg\{
\frac{1}{\epsilon^2}A_{13}^{(-2)}+\frac{1}{\epsilon}A_{13}^{(-1)}+\frac{A_{13}^{(0)}}{\Delta}\bigg\}\ .
\label{pentagon}
\end{align}
Here,
\bqa
\Delta\,&=\,\alpha(\alpha-1)m^4-m^2(t+\alpha(\alpha-1)(t_a+t_b))+\alpha(\alpha-1)(Q^2t+t_at_b)\ ,
\eqa
\begin{align}
A_{13}^{(-2)}\,&=\,\bigg\{-(H_{T}^a+mH_{p\varepsilon}^a)
\big[\frac{1}{t_a-m^2}+\frac{1}{t_b-m^2}\big]
+\frac{sH_{\varepsilon}^a-2(H_{k}^a-mH_m^a)\varepsilon\cdot
p-2H_{p}^a\varepsilon\cdot r}{t_b-m^2}\bigg\}\ ,
\end{align}
and
\begin{align}
A_{13}^{(-1)}\,&=\,2\bigg\{(H_{T}^a
+mH_{p\varepsilon}^a)\big[\ln(m)(\frac{1}{t_a-m^2}
+\frac{1}{t_b-m^2})+\frac{\ln(1-\frac{t_a}{m^2})}{t_a-m^2}
+\frac{\ln(1-\frac{t_b}{m^2})}{t_b-m^2}\nonumber\\
&+\frac{(t_b-t_a)\ln(\frac{\alpha-1}{\alpha})}{(t_a-m^2)(t_b-m^2)}\big]
-\frac{sH_{\varepsilon}^a-2(H_{k}^a-mH_m^a)\varepsilon\cdot
p-2H_{p}^a\varepsilon\cdot
r}{t_b-m^2}\ln\bigg(\frac{(t_a-m^2)\alpha}{m(1-\alpha)}\bigg)\bigg\}\ .
\end{align}
The finite piece $A_{13}^{(0)}$ in (\ref{pentagon}) is listed in the Appendix. Note, we can reproduce the massless result \cite{qiao} when taking the $m\rightarrow0$ limit in above expressions.

We can obtain the amplitude of the conjugate diagram of
Fig.\ref{graphs}.13 by replacing $s\rightarrow -s$ in the
$D_0(1),D_0(2),D_0(3),D_0(4)$ and $D_i(14)\rightarrow -D_i(14)$.
With there replacements one can easily find that the energy
dependence $\text{ln}~s$ terms in Fig.\ref{graphs}.13 cancel the
terms in its conjugate diagram. Therefore, the energy dependence
terms merely come from $A_{2+3}+\bar{A}_{2+3}$.

\section{Renormalization}

The results of our concerned process contain both infrared and ultraviolet
divergences. The ultraviolet divergences may be renormalized via standard procedure, i.e. canceled by counter terms, in modified minimal subtraction ($\overline{MS}$) scheme here. The infrared divergences may be canceled out when the soft gluon radiation process $\gamma^{*}+~reggeon\longrightarrow Q\bar{Q}g$ is taken into account.

The ultraviolet divergences exist only in the self-energy and triangle diagrams, which are
\bqa A_1^{UV}&=& \frac{e e_f}{(4\pi)^{2-\epsilon}}\frac{2(H_{T}^a+m
H_{p\varepsilon}^a)}{t(t_a-m^2)}\Gamma_{qq}^{(0),a}\
\frac{3N_cc_\Gamma}{2\epsilon_{UV}}, \eqa \bqa A_4^{UV}&=& \frac{e
e_f}{(4\pi)^{2-\epsilon}}\frac{2(H_{T}^a+m
H_{p\varepsilon}^a)}{t(t_a-m^2)}\Gamma_{qq}^{(0),a}(-\frac{c_\Gamma}{2N_c\epsilon_{UV}})\
, \eqa \bqa A_5^{UV}&=& \frac{e
e_f}{(4\pi)^{2-\epsilon}}\frac{2(H_{T}^a+m
H_{p\varepsilon}^a)}{t(t_a-m^2)}\Gamma_{qq}^{(0),a}\frac{3N_cc_\Gamma}{2\epsilon_{UV}}\
, \eqa \bqa A_6^{UV}&=& \frac{e
e_f}{(4\pi)^{2-\epsilon}}\frac{2(H_{T}^a+m
H_{p\varepsilon}^a)}{t(t_a-m^2)}\Gamma_{qq}^{(0),a}(-\frac{c_\Gamma}{2N_c\epsilon_{UV}})\
, \eqa \bqa A_{7+8+9}^{UV}&=& \frac{e
e_f}{(4\pi)^{2-\epsilon}}\frac{2(H_{T}^a+m
H_{p\varepsilon}^a)}{t(t_a-m^2)}\Gamma_{qq}^{(0),a}\frac{c_\Gamma}{\epsilon_{UV}}
(\frac{5}{3}N_c-\frac{2}{3}n_f) \ , \eqa \bqa A_{10}^{UV}&=& \frac{e
e_f}{(4\pi)^{2-\epsilon}}\frac{2(H_{T}^a+m
H_{p\varepsilon}^a)}{t(t_a-m^2)}\Gamma_{qq}^{(0),a}\frac{c_\Gamma
}{\epsilon_{UV}}C_F\ , \eqa and \bqa A_{11}^{UV}&=& \frac{e
e_f}{(4\pi)^{2-\epsilon}}\frac{2}{t(t_a-m^2)^2}\Gamma_{qq}^{(0),a}\frac{c_\Gamma
}{\epsilon_{UV}}2C_F((7m^2-t_a)H_{T}^a+2(t_a+2m^2)mH_{p\varepsilon}^a)\
. \eqa

 For the ultraviolet divergence discussed above,  when taking the massless limit we can find it is in agreement with the result in \cite{qiao}. We denote $Z_2, Z_3, Z_m, Z_g$ as the quark-field, gluon-field, mass, and coupling renormalization constants, respectively. Note, in our calculation the renormalization constants $Z_2$ and $Z_m$ are defined in on-shell Scheme, while $Z_3$ and $Z_g$ are given in
$\mathrm{\overline{MS}}$ scheme, which tells:
\bqa Z_2
&=&1-\frac{\alpha_s}{4\pi}C_F\big[\frac{1}{\epsilon_{UV}}+ \frac{2}{\epsilon_{UV}}-3\gamma_E+3\ln(\frac{4\pi
\mu^2}{m^2})+4\big] \ ,
\eqa
\bqa
Z_3 &=&
1+\frac{\alpha_s}{4\pi}(\frac{5}{3}N_c-\frac{2}{3}n_f)\big[
\frac{1}{\epsilon_{UV}}-\gamma_E+\ln(4\pi)\big]\ ,
\eqa
\bqa
Z_g &=&
1-\frac{\alpha_s}{4\pi}(\frac{11}{6}N_c-\frac{1}{3}n_f)\big[
\frac{1}{\epsilon_{UV}}-\gamma_E+\ln(4\pi)\big]\ ,
\eqa
and
\bqa
Z_m &=&1-3\frac{\alpha_s}{4\pi}C_F\big[\frac{1}{\epsilon_{UV}}-\gamma_E+\ln(\frac{4\pi
\mu^2}{m^2})+\frac{4}{3}\big]\ .
\eqa

After including the counter terms, all above ultraviolet divergences and those divergences from their conjugate diagrams, are canceled out. Hence our results will be ultraviolet finite.

\section{Numerical Results}

In the following we show the mass effect numerically in NLO corrections of the  photon impact factor. Since in this work the infrared divergences still exist, the NLO amplitude squared can be expressed as
\begin{eqnarray}
\mathrm{|M_{NLO}|^2}&=&\mathrm{M_{NLO}}\mathrm{M^{*}_{LO}}=(\mathrm{M_{Born}+M_{Loop}})\mathrm{M^{*}_{LO}}\nonumber\\
&=&\mathrm{|M|^2_{Born}}+\frac{1}{\epsilon^2}\mathrm{|M|^2_{Loop~IR2}}+\frac{1}{\epsilon}\mathrm{|M|^2_{Loop~IR1}}
 +\mathrm{|M|^2_{Loop~finite}}\ .
\end{eqnarray}
We then can evaluate respectively the mass effects for the Born term $\mathrm{|M|^2_{Born}}$, second-order infrared divergent term
$\mathrm{|M|^2_{Loop~IR2}}$, first-order infrared divergent term $\mathrm{|M|^2_{Loop~IR1}}$, and NLO finite term
$\mathrm{|M|^2_{Loop~finite}}$.

\begin{figure}
\centering
\includegraphics[width=0.480\textwidth]{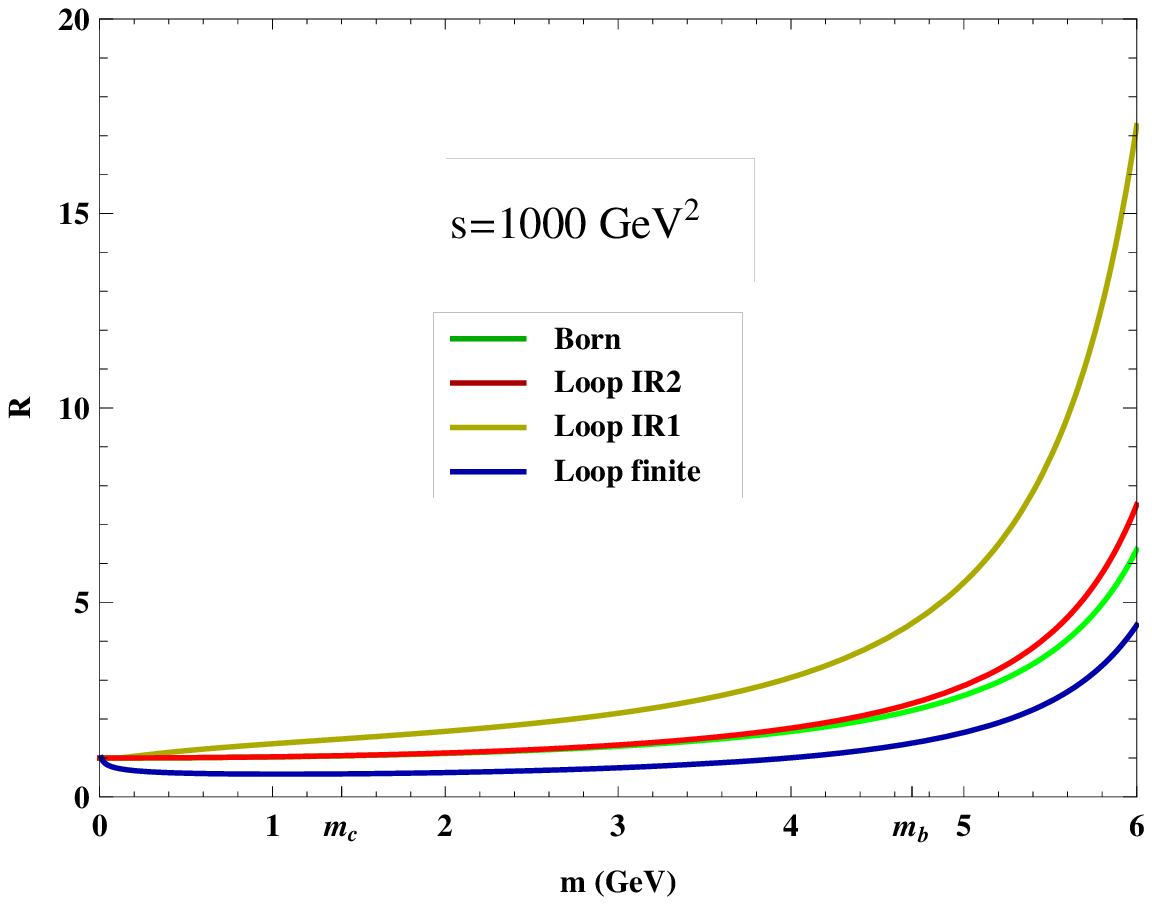}%
\hspace{5mm}
\includegraphics[width=0.480\textwidth]{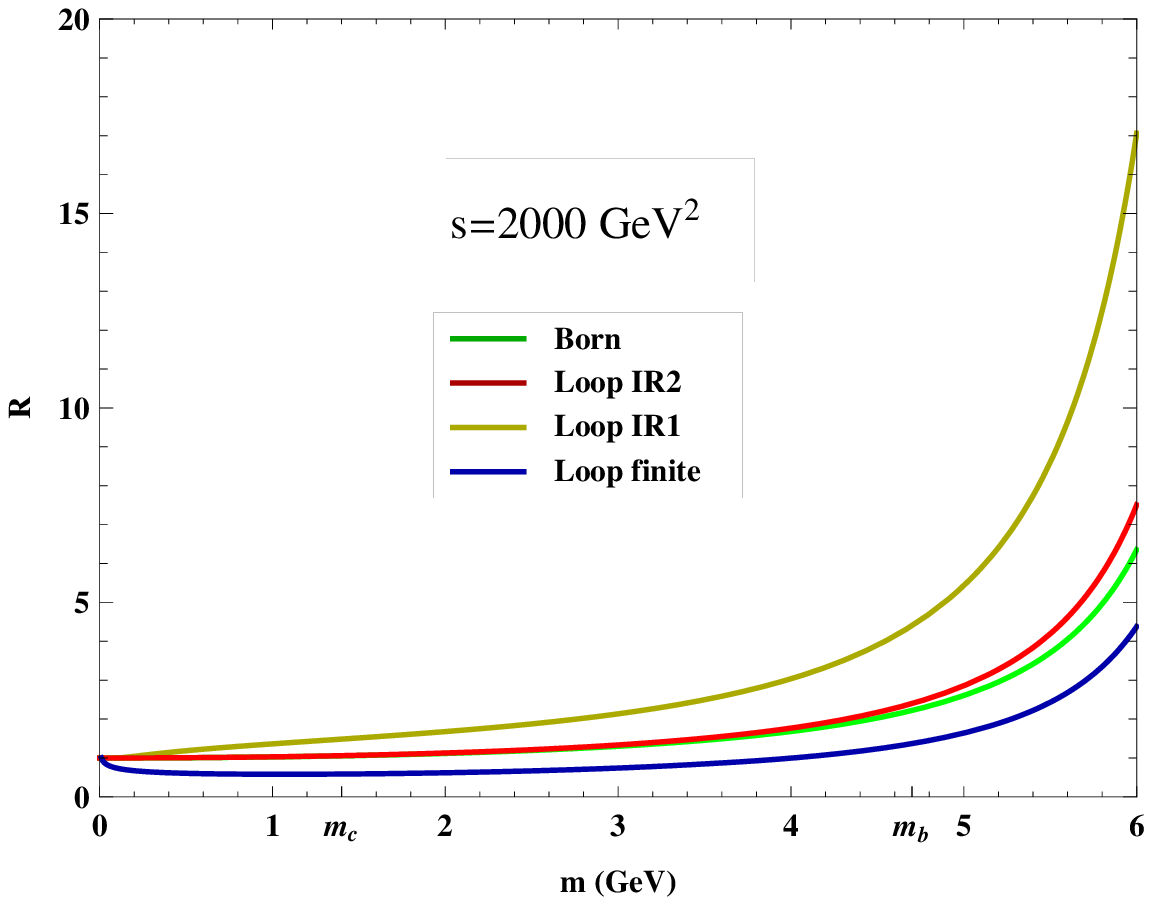}\hspace*{\fill}
\caption{\small the ratios of $\mathrm{|M|^2_{Born}}$,
$\mathrm{|M|^2_{Loop~IR2}}$, $\mathrm{|M|^2_{Loop~IR1}}$ and
$\mathrm{|M|^2_{Loop~finite}}$ versus the quark mass in the process
$\gamma^* + q \to Q\bar{Q} + q$. }
\label{ratio}
\vspace{-0mm}
\end{figure}

In the numerical evaluation, we take the following inputs:
\begin{eqnarray}
n_f=5,\ Q=7~\mathrm{GeV},\ \mu=Q^2,\ \alpha=0.2,\
t_a=t_b=t=50~\mathrm{GeV^2}.
\end{eqnarray}
In our calculation, the quark mass $m$ varies from $0~\mathrm{GeV}$
to $6~\mathrm{GeV}$, then $n_f=5$ (in the physical world the charm
quark mass and bottom quark mass are about 1.4~GeV and 4.7~GeV
respectively). In order to demonstrate the results more clearly, we
define the ratio
$\mathrm{R(m)}=\frac{\mathrm{|M|^2}}{\mathrm{|M|^2}_{m=0~\mathrm{GeV}}}$,
and the ratios of $\mathrm{|M|^2_{Born}}$,
$\mathrm{|M|^2_{Loop~UV}}$, $\mathrm{|M|^2_{Loop~IR2}}$,
$\mathrm{|M|^2_{Loop~IR1}}$ and $\mathrm{|M|^2_{Loop~finite}}$. The
input $s$ are taken as $s=1000~\mathrm{GeV^2}$ and $
s=2000~\mathrm{GeV^2}$, respectively. The curves are shown in
Fig.\ref{ratio}.

From Fig.\ref{ratio} we can see that, for $s=1000~\mathrm{GeV^2}$
and $s=2000~\mathrm{GeV^2}$, the curves change little. As the quark
mass gets larger, the NLO amplitude squared also gets larger
quickly, even for the $\mathrm{|M|^2_{Loop~IR1}}$, the ratio is
nearly $20$ when the quark mass is $6~\mathrm{GeV}$. So we may
conclude that, in the $\gamma^{*} \longrightarrow Q\bar{Q}-Reggeon$
vertex, the quark mass effects on the photon impact factor are
significant, and may influence the results of high energy
photon-photon scattering and heavy quark pair leptoproduction.

\newpage
\section{Conclusions}

In this paper, we calculated the process
$\gamma^{*}+~q\longrightarrow Q\bar{Q}+q$ at NLO with fully virtual corrections in the high energy
limit for massive quark pair $Q(\bar{Q}$, which tells the coupling of the reggeized
gluon to $\gamma^{*}\longrightarrow Q\bar{Q}$.  This calculation is just the first step of our
final purpose, to obtain the complete NLO corrections to the photon impact
factor and checking the BFKL pomeron prediction for the process
$\gamma^{*} + \gamma^{*} \longrightarrow \gamma^{*} + \gamma^{*}$ at
high energies.

In our calculation all contributions from loop diagrams, i.e., self-energy, triangle,
box and pentagon diagrams, are regulated in dimensional
regularization scheme and the ultraviolet divergences are renormalized by
adding the corresponding counter terms. In the end, the infrared divergences still
exist in the result, which would be canceled after taking account of the real corrections.

We find that for $\gamma^{*}+~q\longrightarrow Q\bar{Q}+q$ process in the
massive quark case, the quark mass effects are significant at the next-to-leading order of accuracy, which indicates that
for heavy quark diffractive photoproduction, the quark mass is indispensable. Our result might essentially show the quark mass effect, in spite of the absence of real corrections, which are needed in a complete NLO calculation.
Moreover, for the photon impact factor with
heavy quarks, the photon is legitimate to be real in order to guarantee the perturbative QCD calculations applicable.

\vspace{0.7cm} {\bf Acknowledgments}

This work was supported by the National Natural Science Foundation
of China (NSFC) under grants 11905006, 11805042, 11975236 and 11635009.

\newpage
\appendix{\bf\Large Appendix}

The loop integrals in LoopTools are defined as
\begin{align}
&C_0(p_1^2,p_2^2,p_3^2,(p_1+p_2)^2,m_1^2,m_2^2,m_3^2)
=\frac{(2\pi\mu)^{4-2\epsilon}}{i\pi^2}\int
\frac{d^{4-2\epsilon}q}{[q^2-m_1^2][(q+p_1)^2-m_2^2][(q+p_1+p_2)^2-m_3^2]},\nonumber\\
&~~~~~~~~~~~~~~~~~~~~~~~~~~~~~~~~~~~~~~~~~~~~~~~~~~~~~~~~~~~~~~~~~~~~~~~~~\nonumber\\
&D_0(p_1^2,p_2^2,(p_1+p_2)^2,(p_2+p_3)^2,m_1^2,m_2^2,m_3^2,m_4^2)\nonumber\\
&=\frac{(2\pi\mu)^{4-2\epsilon}}{i\pi^2}\int
\frac{d^{4-2\epsilon}q}{[q^2-m_1^2][(q+p_1)^2-m_2^2][(q+p_1+p_2)^2-m_3^2][(q+p_1+p_2+p_3)^2-m_4^2]},\nonumber\\
&~~~~~~~~~~~~~~~~~~~~~~~~~~~~~~~~~~~~~~~~~~~~~~~~~~~~~~~~~~~~~~~~~~~~~~~~~\nonumber\\
&\frac{(2\pi\mu)^{4-2\epsilon}}{i\pi^2}\int d^{4-2\epsilon}q
\frac{q^{\mu}}{[q^2-m_1^2][(q+p_1)^2-m_2^2][(q+p_1+p_2)^2-m_3^2]}
=\sum_{i=1}^2 C_i p_i^{\mu},\nonumber\\
&~~~~~~~~~~~~~~~~~~~~~~~~~~~~~~~~~~~~~~~~~~~~~~~~~~~~~~~~~~~~~~~~~~~~~~~~~\nonumber\\
&\frac{(2\pi\mu)^{4-2\epsilon}}{i\pi^2}\int d^{4-2\epsilon}q
\frac{q^{\mu}q^{\nu}}{[q^2-m_1^2][(q+p_1)^2-m_2^2][(q+p_1+p_2)^2-m_3^2]}
=g^{\mu\nu}C_{00}+\sum_{i,j=1}^2 C_{ij}
p_i^{\mu}p_j^{\nu},\nonumber\\
&~~~~~~~~~~~~~~~~~~~~~~~~~~~~~~~~~~~~~~~~~~~~~~~~~~~~~~~~~~~~~~~~~~~~~~~~~\nonumber\\
&\frac{(2\pi\mu)^{4-2\epsilon}}{i\pi^2}\int d^{4-2\epsilon}q
\frac{q^{\mu}}{[q^2-m_1^2][(q+p_1)^2-m_2^2][(q+p_1+p_2)^2-m_3^2][(q+p_1+p_2+p_3)^2-m_4^2]}\nonumber\\
&=\sum_{i=1}^3 D_i p_i^{\mu},\nonumber\\
&~~~~~~~~~~~~~~~~~~~~~~~~~~~~~~~~~~~~~~~~~~~~~~~~~~~~~~~~~~~~~~~~~~~~~~~~~\nonumber\\
&\frac{(2\pi\mu)^{4-2\epsilon}}{i\pi^2}\int d^{4-2\epsilon}q
\frac{q^{\mu}q^{\nu}}{[q^2-m_1^2][(q+p_1)^2-m_2^2][(q+p_1+p_2)^2-m_3^2][(q+p_1+p_2+p_3)^2-m_4^2]}\nonumber\\
&=g^{\mu\nu}D_{00}+\sum_{i,j=1}^3 D_{ij} p_i^{\mu}p_j^{\nu},\nonumber\\
&~~~~~~~~~~~~~~~~~~~~~~~~~~~~~~~~~~~~~~~~~~~~~~~~~~~~~~~~~~~~~~~~~~~~~~~~~\nonumber\\
&\frac{(2\pi\mu)^{4-2\epsilon}}{i\pi^2}\int d^{4-2\epsilon}q
\frac{q^{\mu}q^{\nu}q^{\rho}}{[q^2-m_1^2][(q+p_1)^2-m_2^2][(q+p_1+p_2)^2-m_3^2][(q+p_1+p_2+p_3)^2-m_4^2]}\nonumber\\
&=\sum_{i=1}^3(g^{\mu\nu}p_i^{\rho}+g^{\nu\rho}p_i^{\mu}+g^{\mu\rho}p_i^{\nu})D_{00i}+\sum_{i,j,k=1}^3 D_{ijk} p_i^{\mu}p_j^{\nu}p_k^{\rho}~.\nonumber\\
\end{align}

All the coefficients such as $C_{ij}$, $D_{ijk}$ can be evaluated
numerically by LoopTools.

In the following, we list the expressions that are not given in
the main body of the paper:

\begin{align}
A_{14}\,&=\,\Gamma_{qq}^{(0),a}\frac{2s}{t}\bigg(\frac{e
e_f}{s}\bigg)\bigg(\frac{N_c}{2}\bigg)\frac{1}{(4\pi)^2}\bigg\{
H_{T}^a\big[3m^2D_{0}(14)+2(3m^2-t+t_b)D_{1}(14)\nonumber\\
&+(2m^2+Q^2-2t+2t_a+4t_b)D_{2}(14)
-3Q^2D_{3}(14)+6D_{00}(14)+5m^2D_{11}(14)\nonumber\\
&+(4m^2+Q^2-5t+t_a+5t_b)D_{12}(14)
+4(m^2-Q^2-t_a)D_{13}(14)\nonumber\\
&+(m^2+Q^2+4t_b)D_{22}(14)-2(2m^2+Q^2-2t_b)D_{23}(14)
-3Q^2D_{33}(14)\big]\nonumber\\
&+mH_{p\varepsilon}^a\big[(3m^2+Q^2+2t_a
-2t_b)D_{0}(14)+(5m^2+2Q^2-2t+3t_a)D_{1}(14)\nonumber\\
&+(3m^2+3Q^2-2t+4t_a
+t_b)D_{2}(14)-Q^2D_{3}(14)+6D_{00}(14)\nonumber\\
&+(4m^2+Q^2+t_a)D_{11}(14)+
(4m^2+3Q^2-5t+2t_a+4t_b)D_{12}(14)\nonumber\\
&+2(2m^2-Q^2-2t_a)D_{13}(14)+(2m^2+2Q^2+3t_b)D_{22}(14)\nonumber\\
&+4(t_b-m^2)D_{23}(14)-3Q^2D_{33}(14)\big]\nonumber\\
&+H_{p}^a\big[\varepsilon\cdot
r\big(4m^2D_{1}(14)+2(Q^2-3m^2+2t_a)D_{2}(14)-4Q^2D_{3}(14)+8D_{00}(14)\nonumber\\
&+4m^2D_{11}(14)-2(m^2-Q^2-t_a)D_{12}(14)+4(m^2-Q^2-t_a)D_{13}(14)\nonumber\\
&-2(m^2-Q^2-2t+t_b)D_{22}(14)+4(t_a-m^2)D_{23}(14)
-4Q^2D_{33}(14)\nonumber\\
&-20D_{002}(14)-2(m^2-t+t_b)D_{122}(14)
+2(Q^2-m^2+t_a)D_{123}(14)\nonumber\\
&-2t_bD_{222}(14)+2Q^2D_{233}(14)+2(m^2+Q^2-t_b)D_{223}(14)\big)\nonumber\\
&+\varepsilon\cdot
k\big(-6m^2D_{0}(14)-2(5m^2-2t+2t_b)D_{1}(14)-2(5m^2-2t+2t_b)D_{2}(14)\nonumber\\
&-2Q^2D_{3}(14)+4D_{00}(14)+(4t-6m^2-4t_b)D_{11}(14)+4(t_a-t_b)D_{13}(14)\nonumber\\
&-10(m^2-t+t_b)D_{12}(14)-2(2m^2-2t+3t_b)D_{22}(14)+4(t_a-t_b)D_{23}(14)\nonumber\\
&-2Q^2D_{33}(14)-20D_{001}(14)-20D_{002}(14)-2(2m^2-t+t_b)D_{112}(14)\nonumber\\
&+2(Q^2-m^2+t_a)D_{113}(14)-2(m^2-t+2t_b)D_{122}(14)+2Q^2D_{133}(14)\nonumber\\
&+2(2Q^2+t_a-t_b)D_{123}(14)-2t_bD_{222}(14)+2(m^2+Q^2-t_b)D_{233}(14)\nonumber
\end{align}
\begin{align}
&+2Q^2D_{233}(14)\big) \big]+sH_{\varepsilon}^a\big[-2\alpha
m^2D_{0}(14)+(m^2+Q^2-(7m^2+Q^2)\alpha)D_{1}(14)\nonumber\\
&+((3-5\alpha)m^2+Q^2+2t-2t_a-2(Q^2+t_b)\alpha)D_{2}(14)-2(1+2\alpha)D_{00}(14)\nonumber\\
&+(m^2+Q^2(1+2\alpha))D_{3}(14)-((1+3\alpha)m^2
-(1-\alpha)(Q^2+t_a))D_{11}(14)\nonumber\\
&+((2-3\alpha)Q^2+t_a-t_b
-\alpha(3m^2+2t_a+3t_b)+2(\alpha+1)t)D_{12}(14)\nonumber\\
&+((Q^2+t_a)(2\alpha+3)-(1+2\alpha)m^2)D_{13}(14)\nonumber\\
&+(m^2+Q^2-t_b-2\alpha(m^2+Q^2+t_b))D_{22}(14)-2D_{003}(14)\nonumber\\
&+((3+2\alpha)m^2+3Q^2-(1+2\alpha)t_b)D_{23}(14)+2(1+\alpha)Q^2D_{33}(14)\nonumber\\
&+2(\alpha-1)(D_{001}(14)+D_{002}(14))+(\alpha-1)
(2m^2-t+t_b)D_{112}(14)\nonumber\\
&+((\alpha-2)m^2+(1-\alpha)(Q^2+t_a))D_{113}(14)
+(m^2-t+2t_b)(\alpha-1)D_{122}(14)\nonumber\\
&+(-m^2+t+(\alpha-2)t_b-2Q^2(\alpha-1)+(1-\alpha)
t_a)D_{123}(14)\nonumber\\
&+(t_a-m^2+(2-\alpha)Q^2)D_{133}(14)+(\alpha-1)t_bD_{222}(14)\nonumber\\
&+(m^2+Q^2-2t_b-\alpha(m^2+Q^2-t_b))D_{223}(14)\nonumber\\
&+(m^2-t_b-Q^2(\alpha-2))D_{233}(14)+Q^2D_{333}(14)\big]\nonumber\\
&+2mH_{m}^a\big[2s\big((1+\alpha)D_{2}(14)+2D_{12}(14)+(3-\alpha))D_{22}(14)
+(\alpha+2)D_{23}(14)\nonumber\\
&+(1-\alpha)D_{112}(14)+3(1-\alpha)D_{122}(14)+(2-\alpha)D_{123}(14)+2(1-\alpha)
D_{222}(14)\nonumber\\
&+(3-\alpha)D_{223}(14)+2D_{233}(14)\big)\varepsilon\cdot
r+2s\big(2\alpha D_{0}(14)+(1+3\alpha)D_{1}(14)+4\alpha
D_{2}(14)\nonumber\\
&+2D_{11}(14)+4D_{12}(14)+(2+\alpha)D_{13}(14)
+2D_{22}(14)+(1+\alpha)D_{23}(14)+D_{133}(14)\nonumber\\
&+(1-\alpha)D_{111}(14)+4(1-\alpha)D_{112}(14)+(2-\alpha)D_{113}(14)+5(1-\alpha)D_{122}(14)
\nonumber\\
&+(5-2\alpha)D_{123}(14)+2(1-\alpha)D_{222}(14)+(3-\alpha)D_{223}(14)
+D_{233}(14) \big)\varepsilon\cdot
k\nonumber\\
&+\big(-(m^2+Q^2+2t_a)D_{0}(14)-(2m^2+2Q^2+3t_a)D_{1}(14)
-(m^2+3Q^2-2t\nonumber\\
&+4t_a+t_b)D_{2}(14)-(m^2+Q^2)D_{3}(14)
-6D_{00}(14)-(Q^2+t_a)D_{11}(14)\nonumber\\
&-(3Q^2-t+2t_a)D_{12}(14)-2Q^2D_{13}(14)+(t_b-2m^2-2Q^2)D_{22}(14)\nonumber\\
&-4Q^2D_{23}(14)-Q^2D_{33}(14)+2D_{001}(14)
+4D_{002}(14)+2D_{003}(14)+m^2D_{111}(14)\nonumber\\
&+(3m^2-t+t_b)D_{112}(14)+(2m^2-Q^2-t_a)D_{113}(14)
+(2m^2-2t+3t_b)D_{122}(14)\nonumber\\
&+(2m^2-3Q^2-t-2t_a+2t_b)D_{123}(14)
+(m^2-2Q^2-t_a)D_{133}(14)+2t_bD_{222}(14)\nonumber\\
&+(3t_b-2m^2-2Q^2)D_{223}(14)+(t_b-m^2
-3Q^2)D_{233}(14)-Q^2D_{333}(14)\big)\varepsilon\cdot
p\big]\nonumber\\
&+2H_k^a\big[2s\big((1-\alpha)D_{2}(14)+(1-\alpha)D_{12}(14)-\alpha
D_{23}(14)+(\alpha-1)D_{122}(14)\nonumber\\
&+(\alpha-1)D_{123}(14)+(\alpha-1)D_{222}(14)
+(\alpha-2)D_{223}(14)-D_{233}(14)\big)\varepsilon\cdot r\nonumber
\end{align}
\begin{align}
&+2s\big((1-\alpha)D_{2}(14)-(1+\alpha)D_{13}(14)-\alpha
D_{23}(14)+(\alpha-1)D_{112}(14)\nonumber\\
&+(\alpha-1)D_{113}(14)+2(\alpha-1)D_{122}(14)
+(2\alpha-3)D_{123}(14)-D_{133}(14)\nonumber\\
&+(\alpha-1)D_{222}(14)+(\alpha-2)D_{223}(14)-D_{233}(14)
\big)\varepsilon\cdot k+\big(2m^2D_{0}(14)\nonumber\\
&+4m^2D_{1}(14)+(m^2+Q^2+2(t_a+t_b-t))D_{2}(14)+(m^2-Q^2)D_{3}(14)\nonumber\\
&+2D_{00}(14)+(m^2+Q^2-2t+t_a+2t_b)D_{12}(14)
+(m^2-Q^2-t_a)D_{13}(14)\nonumber\\
&+(m^2+Q^2+t_b)D_{22}(14)+(Q^2-m^2+t_b)D_{23}(14)
-2D_{002}(14)-2D_{003}(14)\nonumber\\
&-m^2D_{112}(14)-m^2D_{113}(14)+(t-m^2-t_b)D_{122}(14)+(Q^2-2m^2+t+t_a\nonumber\\
&-t_b)D_{123}(14)+(Q^2-m^2+t_a)D_{133}(14)
-t_bD_{222}(14)+(m^2+Q^2-2t_b)D_{223}(14)\nonumber\\
&+2m^2D_{11}(14)+(m^2+2Q^2-t_b)D_{233}(14)+Q^2D_{333}(14)\big)\varepsilon\cdot
p\big]\nonumber\\
&-4mH_{pk}^a\big[\big(2(D_{0}(14)+D_{11}(14)+D_{22}(14))+4(D_{1}(14)+D_{2}(14)+D_{12}(14))\big)\varepsilon\cdot
k\nonumber\\
&+\big(D_{0}(14)+D_{1}(14)+3D_{2}(14)+2D_{12}(14)+2D_{22}(14)\big)\varepsilon\cdot
r\big]\nonumber\\
&+2msH_{k\varepsilon}^a\big((1-2\alpha)D_{0}(14)+(2-3\alpha)D_{1}(14)
+(2-3\alpha)D_{2}(14)+(1-\alpha)D_{11}(14)\nonumber\\
&+2(1-\alpha)D_{12}(14)
+(1-\alpha)D_{22}(14)+D_{13}(14)+D_{23}(14)\big)\bigg\}.
\end{align}
and $D_i(14)=D_i(m^2,t,m^2,-Q^2,t_b,t_a,m^2,0,0,m^2)$.

\begin{align}
A_{12}^{0}\,&=\,
-H_{T}^a\big[(3m^2D_{0}(12)_{fin}+(4m^2-M^2)D_{1}(12)_{fin}+m^2D_{11}(12)_{fin})\nonumber\\
&+(2m^2+M^2)(D_{2}(12)+D_{12}(12)+D_{22}(12)-D_{123}(12)-D_{223}(12))\nonumber\\
&-2D_{00}(12)+Q^2D_{33}(12)-6D_{003}(12)+Q^2D_{333}(12)\nonumber\\
&-(m^2-Q^2-t_a)D_{133}(12)
-(2m^2-2Q^2-t_a-t_b)D_{233}(12)\big]\nonumber\\
&+mH_{p\varepsilon}^a\big[-(3m^2+t_a-t_b)D_{0}(12)_{fin}
-(4m^2+2Q^2-t+3t_a)D_{1}(12)_{fin}\nonumber\\
&-(Q^2+t_a)D_{11}(12)_{fin}+m^2D_{111}(12)_{fin}-(2m^2+M^2)(D_{2}(12)+D_{22}(12)\nonumber\\
&-D_{112}(12)-D_{122}(12)-D_{223}(12))-(Q^2+t)D_{12}(12)-2Q^2D_{13}(12)\nonumber\\
&-Q^2D_{33}(12) +6D_{001}(12)+6D_{003}(12)+(2m^2-Q^2-t_a)D_{113}(12)\nonumber\\
& +(4m^2-3Q^2+t-2t_a-2t_b)D_{123}(12)
+(m^2-2Q^2-t_a)D_{133}(12)\nonumber\\
&+(2m^2-2Q^2-t_a-t_b)D_{233}(12)-Q^2D_{333}(12)+2D_{00}(12)\big]\nonumber\\
& +sH_{\varepsilon}^a\big[\alpha
m^2(D_{0}(12)_{fin}+3D_{1}(12)_{fin}+2D_{11}(12)_{fin})+2\alpha
D_{00}(12)\nonumber\\
&+((1+\alpha)m^2+t-t_a-\alpha(2Q^2+t_a+t_b))D_{2}(12)-\alpha Q^2
D_{3}(12)\nonumber\\
&+(4\alpha m^2+(1+\alpha)t-(1+2\alpha)t_a-2\alpha t_b-3\alpha
Q^2)D_{12}(12)\nonumber\\
&-(2\alpha Q^2-\alpha m^2+(1+\alpha)t_a)D_{13}(12)+\alpha(2m^2-2Q^2-t_a-t_b)D_{223}(12)\nonumber\\
&+((1+4\alpha)m^2+(1+\alpha)t-3\alpha Q^2-(1+2\alpha)t_a-2\alpha
t_b)D_{22}(12)\nonumber\\
&+((1+2\alpha)m^2-(1+\alpha)t_a-\alpha
t_b-4\alpha Q^2)D_{23}(12)-\alpha Q^2D_{233}(12)\nonumber\\
&-\alpha Q^2D_{33}(12)+2(\alpha-1)D_{001}(12)
-2(1-3\alpha)D_{002}(12)-2D_{003}(12)\nonumber\\
&+\alpha(2m^2+M^2)(D_{122}(12)+D_{222}(12))
+\alpha(m^2-Q^2-t_a)D_{123}(12)\big]\nonumber\\
&+sH_{k\varepsilon}^a\big[(2\alpha-1)D_{0}(12)_{fin}
+(3\alpha-2)D_{1}(12)_{fin}\nonumber\\
&+(\alpha-1)D_{11}(12)_{fin}-D_{12}(12)-D_{13}(12)\big]\nonumber\\
&+2mH_{pk}^a\big[(2D_{0}(12)_{fin}+3D_{1}(12)_{fin}
+D_{11}(12)_{fin})\varepsilon\cdot k\nonumber\\
&+(D_{0}(12)_{fin}+D_{1}(12)_{fin}-D_{12}(12))\varepsilon\cdot r\big]\nonumber\\
&+2sH_k^a\big[(D_{2}(12)+D_{12}(12)+(1+\alpha)D_{13}(12)+D_{22}(12)+D_{23}(12)
\nonumber\\
&+(1-\alpha)D_{113}(12)+D_{123}(12)+D_{133}(12))\varepsilon\cdot
k+((1-\alpha)(D_{2}(12)\nonumber\\
&+D_{12}(12)+D_{22}(12)-D_{123}(12))-\alpha
D_{23}(12)-D_{223}(12)-D_{233}(12))\varepsilon\cdot r\big]\nonumber
\end{align}
\begin{align}
~~~~&+2msH_{m}^a\big[(-2\alpha
D_{0}(12)_{fin}-(1+3\alpha)D_{1}(12)_{fin}-2D_{11}(12)_{fin}
+(\alpha-1)D_{111}(12)_{fin}\nonumber\\
&-D_{2}(12)-2D_{12}(12)-(\alpha+2)D_{13}(12)-D_{22}(12)-D_{23}(12)
-D_{112}(12)\nonumber\\
&+(\alpha-2)D_{113}(12)-D_{123}(12))\varepsilon\cdot k
+((1+\alpha)D_{2}(12)+2D_{12}(12)+(1+\alpha)D_{22}(12)\nonumber\\
&+(2+\alpha)D_{23}(12)
+(1-\alpha)D_{112}(12)+(2-\alpha)D_{123}(12)+D_{122}(12)+D_{223}(12)\nonumber\\
&+D_{233}(12))\varepsilon\cdot r \big]\nonumber\\
&-2H_{p}^a\big[(-2m^2D_{0}(12)_{fin}-(2m^2-M^2)D_{1}(12)_{fin}+(m^2+M^2)D_{11}(12)_{fin}
\nonumber\\
&-(m^2+Q^2+t)D_{2}(12)-(m^2+Q^2)D_{3}(12)
+(4m^2-3Q^2+t-2t_a-2t_b)D_{12}(12)\nonumber\\
&+4D_{00}(12)+(2m^2-3Q^2-t_a-t_b)D_{13}(12)
+(2m^2-2Q^2-t_a-t_b)D_{22}(22)\nonumber\\
&+(2m^2-4Q^2-t_a-t_b)D_{23}(12)-2Q^2D_{33}(12)
+6D_{002}(12)+6D_{003}(12)\nonumber\\
&+(3m^2+M^2)D_{112}(12)
+(2m^2-Q^2-t_a)D_{113}(12)+2(2m^2+M^2)D_{122}(12)\nonumber\\
&+(5m^2-4Q^2+t-3t_a-2t_b)D_{123}(12)+m^2D_{111}(12)_{fin}+(2m^2+M^2)D_{222}(12)\nonumber\\
&+(m^2-2Q^2-t_a)D_{133}(12)
+(4m^2-3Q^2+t-2t_a-2t_b)D_{223}(12)+4D_{001}(12)\nonumber\\
&+(2m^2-3Q^2-t_a-t_b)D_{233}(12)-Q^2D_{333}(12))\varepsilon\cdot
k+(m^2D_{1}(12)_{fin}+m^2D_{11}(12)_{fin}\nonumber\\
&+(Q^2-m^2+t_a)D_{2}(12)-Q^2D_{3}(12)+2D_{00}(12)-Q^2D_{33}(12)+2D_{002}(12)\nonumber\\
&+(m^2-Q^2-t_a)(D_{13}(12)+D_{22}(12))+(m^2-2Q^2-t_a)D_{23}(12))\varepsilon\cdot r\big]\nonumber\\
&+2\varepsilon\cdot p
H_{k}^a\big[-m^2D_{0}(12)_{fin}-2m^2D_{1}(12)_{fin}
-m^2D_{11}(12)_{fin}-(2m^2+M^2)\times\nonumber\\
&(D_{2}(12)+D_{12}(12)+D_{22}(12)-D_{123}(12)-D_{223}(12))
-m^2D_{3}(12)-Q^2D_{33}(12)\nonumber\\
&+4D_{003}(12) +m^2D_{113}(12)+(2m^2-2Q^2
-t_a-t_b)D_{233}(12)-Q^2D_{333}(12)\nonumber\\
&+(m^2-Q^2-t_a)D_{133}(12)\big]\nonumber\\
&+2mH_{m}^a\varepsilon\cdot
p\big[(m^2+Q^2+t_a)D_{0}(12)_{fin}+(m^2+2Q^2+2t_a)D_{1}(12)_{fin}
\nonumber\\
&+(Q^2-m^2+t_a)D_{11}(12)_{fin}-m^2D_{111}(12)_{fin}+(Q^2+t)D_{2}(12)\nonumber\\
&+(m^2+2Q^2)D_{3}(12)
+(2Q^2-2m^2+t_a+t_b)D_{12}(12)+2Q^2D_{33}(12)\nonumber\\
&-(m^2-3Q^2-t_a)D_{13}(12)-(2m^2-2Q^2-t_a-t_b)D_{23}(12)
-4D_{001}(12)\nonumber\\
&-(2m^2+M^2)(D_{112}(12)+D_{122}(12)+D_{223}(12))
+(Q^2-2m^2+t_a)D_{113}(12)\nonumber\\
&-(4m^2-3Q^2+t-2t_a-2t_b)D_{123}(12)-(m^2-2Q^2-t_a)D_{133}(12)\nonumber\\
&-4D_{003}(12)+(2Q^2-2m^2+t_a+t_b)D_{233}(12)+Q^2D_{333}(12)\big].
\end{align}
here $D_i(12)=D_i(m^2,m^2,t,-Q^2,2m^2+M^2,t_a,m^2,0,m^2,m^2)$.

\begin{align}
\bar{A}_{12}^{0}\,&=\,
-H_{T}^a\big[(3m^2D_{0}(\bar{12})_{fin}+(4m^2-M^2)D_{1}(\bar{12})_{fin}
+m^2D_{11}(\bar{12})_{fin})\nonumber\\
&+(2m^2+M^2)(D_{2}(\bar{12})+D_{12}(\bar{12})
+D_{22}(\bar{12})-D_{123}(\bar{12})-D_{223}(\bar{12}))\nonumber\\
&-2D_{00}(\bar{12})+Q^2D_{33}(\bar{12})-6D_{003}(\bar{12})
-(m^2-Q^2-t_b)D_{133}(\bar{12})\nonumber\\
&-(2m^2-2Q^2-t_b-t_a)D_{233}(\bar{12})+Q^2D_{333}(\bar{12})\big]\nonumber\\
&-mH_{p\varepsilon}^a\big[(3m^2+t_a-t_b)D_{0}(\bar{12})_{fin}
+(4m^2-t+2t_a-t_b)D_{1}(\bar{12})_{fin}\nonumber\\
&+(2m^2-Q^2-t_b)D_{11}(\bar{12})_{fin}+m^2D_{111}(\bar{12})_{fin}
+m^2D_{3}(\bar{12})\nonumber\\
&+(2m^2+M^2)(D_{2}(12)+D_{22}(12)+D_{112}(\bar{12})
+D_{122}(\bar{12})-D_{223}(\bar{12}))\nonumber\\
&+(4m^2-3Q^2+t-2t_a-2t_b)D_{12}(\bar{12})
-2Q^2D_{13}(\bar{12})+Q^2D_{33}(\bar{12})\nonumber\\
&+6D_{001}(\bar{12})-6D_{003}(\bar{12})-(Q^2+t_b)D_{113}(\bar{12})
-(Q^2+t)D_{123}(\bar{12})\nonumber\\
&+(m^2-t_b)D_{133}(\bar{12})-(2m^2-2Q^2-t_a-t_b)D_{233}(\bar{12})
+Q^2D_{333}(\bar{12})\big]\nonumber\\
&+sH_{\varepsilon}^a\big[(\alpha+2)
m^2D_{0}(\bar{12})_{fin}+((3\alpha+1)m^2-M^2)D_{1}(\bar{12})_{fin}\nonumber\\
&+(2\alpha-1)m^2D_{11}(\bar{12})_{fin}+(\alpha m^2
+(1-2\alpha)Q^2-\alpha
t_a+(1-\alpha)t_b)D_{2}(\bar{12})\nonumber\\
&+(m^2+(1-\alpha)Q^2)D_{3}(\bar{12})+2(\alpha-2)
D_{00}(\bar{12})\nonumber\\
&+(2(2\alpha-1)m^2+(2-3\alpha)Q^2+(\alpha-1)(t+2t_b)+(1-2\alpha)t_a)D_{12}(\bar{12})\nonumber\\
&+(2(1-\alpha)Q^2+(\alpha-1)m^2+(2-\alpha)t_b)D_{13}(\bar{12})\nonumber\\
&+((4\alpha-3)m^2+(\alpha-1)t+(2-3\alpha)
Q^2+(1-2\alpha)t_a+2(1-\alpha)t_b)D_{22}(\bar{12})\nonumber\\
&+((2\alpha-3)m^2+(1-\alpha)t_a+(2-\alpha)
t_b-4(1-\alpha)Q^2)D_{23}(\bar{12})\nonumber\\
&+(2-\alpha)Q^2 D_{33}(\bar{12})+2\alpha
D_{001}(\bar{12})+2(3\alpha-2)D_{002}(\bar{12})
-4D_{003}(\bar{12})\nonumber\\
&+(\alpha-1)(2m^2+M^2)(D_{122}(\bar{12})
+D_{222}(\bar{12}))+(Q^2-m^2+t_b)D_{133}(\bar{12})\nonumber\\
&+(2(\alpha-2)m^2+(3-2\alpha)Q^2-t+(2-\alpha)(t_a+t_b))D_{223}(\bar{12})\nonumber\\
&+((\alpha-3)m^2+(2-\alpha)Q^2-t+t_a+(2-\alpha)t_b)D_{123}(\bar{12})
+(\alpha-1)m^2D_{112}(\bar{12})\nonumber\\
&+((3-\alpha)Q^2-2m^2+t_a+t_b)D_{233}(\bar{12})+
Q^2D_{333}(\bar{12})-m^2D_{113}(\bar{12})\big]\nonumber\\
&+sH_{k\varepsilon}^a\big[(2\alpha-1)D_{0}(\bar{12})_{fin}
+(3\alpha-1)D_{1}(\bar{12})_{fin}+\alpha D_{11}(\bar{12})_{fin}
+D_{12}(\bar{12})\nonumber
\end{align}
\begin{align}
&+D_{13}(\bar{12})\big]+2mH_{pk}^a\big[(2D_{0}(\bar{12})_{fin}+3D_{1}(\bar{12})_{fin}
+D_{11}(\bar{12})_{fin})\varepsilon\cdot k\nonumber\\
&+(D_{0}(\bar{12})_{fin}+2D_{1}(\bar{12})_{fin}+D_{11}(\bar{12})_{fin}
+D_{12}(\bar{12}))\varepsilon\cdot r\big]\nonumber\\
~~~~&+2sH_k^a\big[(D_{2}(\bar{12})+D_{12}(\bar{12})+(2-\alpha)D_{13}(\bar{12})
+D_{22}(\bar{12})\nonumber\\
&+D_{23}(\bar{12})+\alpha D_{113}(\bar{12})+D_{123}(\bar{12})
+D_{133}(\bar{12}))\varepsilon\cdot k\nonumber\\
&+((1-\alpha)(D_{2}(\bar{12})+D_{12}(\bar{12})+D_{22}(\bar{12}))+\alpha D_{113}(\bar{12})+D_{133}(\bar{12})\nonumber\\
&+(2-\alpha)(D_{13}(\bar{12})+D_{23}(\bar{12}))
+(1+\alpha)D_{123}(\bar{12})+D_{223}(\bar{12})+D_{233}(\bar{12}))\varepsilon\cdot
r\big]\nonumber\\
&+2msH_{m}^a\big[(-2\alpha
D_{0}(\bar{12})_{fin}+(1-3\alpha)D_{1}(\bar{12})_{fin}+D_{11}(\bar{12})_{fin}
+\alpha D_{111}(\bar{12})_{fin}\nonumber\\
&-D_{2}(\bar{12})-(1-\alpha)D_{13}(\bar{12})-D_{22}(\bar{12})-D_{23}(\bar{12})
+D_{112}(\bar{12})+(1-\alpha)D_{113}(\bar{12})\nonumber\\
&-D_{123}(\bar{12})-D_{133}(\bar{12}))\varepsilon\cdot
k+(D_{1}(\bar{12})_{fin}+(1+\alpha)D_{11}(\bar{12})_{fin}+\alpha D_{111}(\bar{12})_{fin}\nonumber\\
&-D_{2}(\bar{12})+(1+\alpha)(2D_{12}(\bar{12})+D_{22}(\bar{12})+D_{112}(\bar{12}))
+(1-\alpha)D_{113}(\bar{12})\nonumber\\
&+\alpha D_{23}(\bar{12}) -\alpha
D_{123}(\bar{12})+D_{122}(\bar{12})-D_{133}(\bar{12})
-D_{223}(\bar{12})-D_{233}(\bar{12}))\varepsilon\cdot r \big]\nonumber\\
&-2H_{p}^a\big[(-2m^2D_{0}(\bar{12})_{fin}
-(2m^2-M^2)D_{1}(\bar{12})_{fin}+(m^2+M^2)D_{11}(\bar{12})_{fin}
\nonumber\\
&+m^2D_{111}(\bar{12})_{fin}-(m^2+Q^2+t)D_{2}(\bar{12})-(m^2+Q^2)D_{3}(\bar{12})+4D_{00}(\bar{12})\nonumber\\
&+(4m^2-3Q^2+t-2t_a-2t_b)D_{12}(\bar{12})+(2m^2-3Q^2-t_a-t_b)D_{13}(\bar{12})\nonumber\\
&+4D_{001}(\bar{12})
+(2m^2-2Q^2-t_a-t_b)D_{22}(\bar{12})+6D_{002}(\bar{12})+6D_{003}(\bar{12})\nonumber\\
&+(2m^2-4Q^2-t_a-t_b)D_{23}(\bar{12})-2Q^2D_{33}(\bar{12})
+(3m^2+M^2)D_{112}(\bar{12})\nonumber\\
& +(2m^2-Q^2-t_b)D_{113}(\bar{12})
+(5m^2-4Q^2+t-2t_a-3t_b)D_{123}(\bar{12})\nonumber\\
&+(m^2-2Q^2-t_b)D_{133}(\bar{12})
+(4m^2-3Q^2+t-2t_a-2t_b)D_{223}(\bar{12})\nonumber\\
&+2(2m^2+M^2)D_{122}(\bar{12})+(2m^2+M^2)D_{222}(\bar{12})
-Q^2D_{333}(\bar{12})\nonumber\\
&+(2m^2-3Q^2-t_a-t_b)D_{233}(\bar{12})
)\varepsilon\cdot k\nonumber\\
&+(m^2D_{0}(\bar{12})_{fin}+m^2D_{1}(\bar{12})_{fin}+m^2D_{11}(\bar{12})_{fin}+m^2D_{111}(\bar{12})_{fin}\nonumber\\
&-(Q^2+t_a)D_{2}(\bar{12})+(4m^2-3Q^2+2t-3t_a-2t_b)D_{12}(\bar{12})\nonumber\\
&+(m^2-2Q^2-t_a)D_{13}(\bar{12})+(3m^2-2Q^2+t-2t_a-t_b)D_{22}(\bar{12})\nonumber\\
&+(m^2-2Q^2-t_a)D_{23}(\bar{12})+4D_{001}(\bar{12})
+(3m^2+M^2)D_{112}(\bar{12})\nonumber
\end{align}
\begin{align}
&+4D_{002}(\bar{12})+(m^2-Q^2-t_b)D_{113}(\bar{12})+2(2m^2+M^2)D_{122}(\bar{12})\nonumber\\
&+(3m^2-3Q^2-t_a-2t_b)D_{123}(\bar{12})
+(2m^2+M^2)D_{222}(\bar{12})\nonumber\\
&+(2m^2-2Q^2-t_a-t_b)D_{223}(\bar{12})-Q^2D_{133}(\bar{12})-Q^2D_{233}(\bar{12}))\varepsilon\cdot
r\big]\nonumber\\
~~~~&+2\varepsilon\cdot p
H_{k}^a\big[-2m^2D_{0}(\bar{12})_{fin}-(2m^2-M^2)D_{1}(\bar{12})_{fin}
+2D_{00}(\bar{12})_{fin}+2D_{003}(\bar{12})_{fin}\big]\nonumber\\
&+2mH_{m}^a\varepsilon\cdot
p\big[(2m^2+Q^2+t_a)D_{0}(\bar{12})_{fin}+(m^2+2Q^2-t+2t_a+t_b)D_{1}(\bar{12})_{fin}
\nonumber\\
&-m^2D_{11}(\bar{12})_{fin}-(2m^2-2Q^2-t_a-t_b)D_{2}(\bar{12})
+2Q^2D_{3}(\bar{12})-2D_{00}(\bar{12})\nonumber\\
&+(Q^2-2m^2-t+t_a+t_b)D_{12}(\bar{12})-(2m^2-2Q^2-t_a-t_b)D_{23}(\bar{12})\nonumber\\
&+(Q^2-m^2+t_b)D_{13}(\bar{12})+Q^2D_{33}(\bar{12})+2D_{001}(\bar{12})-2D_{003}(\bar{12})\big].
\end{align}
here $D_i(\bar{12})=D_i(m^2,m^2,-Q^2,t,2m^2+M^2,t_b,m^2,0,m^2,m^2)$.

\begin{align}
A_{13}^{(0)}\,&=\,H_{T}^a\big\{\alpha(\alpha-1)t(t_a-t_b)
[(\alpha-1)sD_0(2)-\alpha sD_0(1)]\nonumber\\
&-(\alpha-1)(2\Delta+(\alpha-1)\alpha(t_a-t_b)(t_a-m^2)) sD_0(3)\nonumber\\
&-\alpha (2\Delta+(\alpha-1)\alpha(t_b-t_a)(t_b-m^2))sD_0(4)\nonumber\\
&-[2(\alpha-1)\alpha m^6+m^4(2(\alpha^2-\alpha-1)t
-3(\alpha-1)\alpha(t_a+t_b))\nonumber\\
&+m^2(2t(\alpha(\alpha-1)Q^2-\alpha(\alpha-2)t_a+(1-\alpha^2)t_b)-2t^2\nonumber\\
&+(\alpha-1)\alpha(t_a^2+4t_at_b+t_b^2))+(\alpha-1)\alpha(2t-t_a-t_b)(Q^2t+t_at_b)]
D_0(14)\big\}\nonumber\\
&+(sH_{\varepsilon}^a-2(H_{k}^a-mH_{m})\varepsilon\cdot
p)\big\{t(\alpha-1)\alpha(t_a-m^2)(\alpha
sD_0(1)-(\alpha-1)sD_0(2))\nonumber\\
&-\alpha (-\Delta+m^2t-(\alpha-1)\alpha Q^2t)sD_0(4)+(\alpha-1)^2\alpha(t_a-m^2)^2 sD_0(3)\nonumber\\
&-[-(\alpha-1)\alpha m^6+\alpha m^4((\alpha-1)(2t_a+t_b)-2(\alpha-2)t)\nonumber\\
&+m^2(\alpha
t(-(\alpha-1)Q^2+2\alpha(t_a+t_b)-2(2t_a+t_b))+2t^2-(\alpha-1)\alpha
t_a(t_a+2t_b))\nonumber
\end{align}
\begin{align}
&-(\alpha-1)\alpha(2t-t_a)(Q^2t+t_at_b)]
D_0(14)\big\}\nonumber\\
&+2t\Delta(H_T^a+mH_{p\varepsilon}^a-sH_{\varepsilon}^a-2H_p^a\varepsilon\cdot
k+2(H_k^a-mH_{m}^a)\varepsilon\cdot p)D_0(14)\nonumber\\
&+mH_{p\varepsilon}^a\big\{t(t-(\alpha-1)\alpha(t_a-t_b))(\alpha
sD_0(1)-(\alpha-1)sD_0(2))\nonumber\\
&-[2(\alpha-1)\alpha
m^4-m^2(3t+(\alpha-1)\alpha(t_a+3t_b))\nonumber\\
&+t(2(\alpha-1)\alpha Q^2+t_b)+(\alpha-1)\alpha t_b(t_a+t_b)]\alpha sD_0(4)\nonumber\\
&-[2(\alpha-1)\alpha m^4-m^2(t+(\alpha-1)
\alpha(t_b+3t_a))\nonumber\\
&+t(2(\alpha-1)\alpha
Q^2-t_a)+(\alpha-1)\alpha t_a(t_a+t_b)](\alpha-1) sD_0(3)\nonumber\\
&-[2(\alpha-1)\alpha
m^6+m^4(t(2\alpha^2-4\alpha-1)-3(\alpha-1)\alpha(t_a+t_b))
\nonumber\\
&+m^2(t(-2\alpha^2(t_a+t_b-Q^2)+2\alpha(3t_a+t_b-Q^2)-t_a+t_b)\nonumber\\
&-2t^2+(\alpha-1)\alpha(t_a^2+4t_at_b+t_b^2))\nonumber\\
&+(Q^2t+t_at_b)((2\alpha^2-4\alpha+1)t
-(\alpha-1)\alpha(t_a+t_b))]D_0(14)
\big\}\nonumber\\
&+2mtH_{pk}^a\varepsilon\cdot p\big\{-\alpha tD_0(1)+(\alpha-1)t D_0(2)\nonumber\\
&+\alpha(t_b-m^2)D_0(4)-(\alpha-1)(t_a-m^2)D_0(3)\nonumber\\
&+(1-2\alpha)(m^4-m^2(t_a+t_b)+Q^2t+t_at_b)
D_0(14)/s\big\}\nonumber\\
~~~&+2H_{p}^a\big\{t(\alpha(\alpha-1)s(m^2-t_a)\varepsilon\cdot r-t
m^2\varepsilon\cdot p)\alpha D_0(1)\nonumber\\
&+t(\alpha(\alpha-1) s(t_a-m^2)\varepsilon\cdot r+t
m^2\varepsilon\cdot p)(\alpha-1)D_0(2)\nonumber\\
&+(t_a-m^2)(\alpha(\alpha-1)s(m^2-t_a)\varepsilon\cdot r
-tm^2\varepsilon\cdot p)(\alpha-1)D_0(3)\nonumber\\
&-(s(\Delta-m^2t+(\alpha-1)\alpha Q^2t)\varepsilon\cdot
r+m^2t(m^2-t_b)\varepsilon\cdot p)\alpha D_0(4)\nonumber\\
&-[(2\alpha-1)m^2t(m^4-m^2(t_a+t_b)+Q^2t+t_at_b)\varepsilon\cdot p/s\nonumber\\
&+(m^2-t_a)(\Delta+m^2t(1-2\alpha))\varepsilon\cdot
r-2t\Delta\varepsilon\cdot
k]D_0(14)\nonumber\\
&-2t\Delta\varepsilon\cdot k(D_1(14)
+D_2(14))-2t\Delta\varepsilon\cdot rD_2(14) \big\}.
\end{align}

and
\bqa
D_0(1)&=&D_0(m^2,0,0,t_a,\alpha s,t,m^2,0,0,0)_{fin}\nonumber\\
D_0(2)&=&D_0(m^2,0,0,t_b,(\alpha-1) s,t,m^2,0,0,0)_{fin}\nonumber\\
D_0(3)&=&D_0(m^2,0,\alpha s,-Q^2,(\alpha-1) s,t_a,m^2,0,0,m^2)_{fin}\nonumber\\
D_0(4)&=&D_0(m^2,0,(\alpha-1) s,-Q^2,\alpha
s,t_b,m^2,0,0,m^2)_{fin}. \eqa $D_i(14)$ is defined in the above
paragraphs. The finite parts of these four-point integrals can also
be obtained from \cite{ellis}, or evaluated by LoopTools
numerically.



\newpage
\begin{thebibliography}{99}
\bibitem{pom1} E. Levin, hep-ph/9808486.
\bibitem{pom2} P. V. Landshoff, hep-ph/0108156.
\bibitem{pom3} V. A. Petrov, and A. V. Prokudin, Eur. Phys. J. \textbf{C} \textbf{23} (2002) 135 .
\bibitem{pom4} V. S. Fadin, R. Fiore and A. Quartarolo, Phys. Rev. \textbf{D50} (1994) 2265.
\bibitem{pom5} V. S. Fadin, D. Ivanov, and M. Kotsky, hep-ph/0007119.
\bibitem{bfkl1} F. Yuan and K. T. Chao, Phys. Rev. \textbf{D60} (1999) 094012.
\bibitem{bfkl2} F. Yuan and K. T. Chao, Phys. Rev. \textbf{D58} (1998) 114016.
\bibitem{bfkl3} V. S. Fadin, and L. N. Lipatov, Phys. Lett. \textbf{B} \textbf{429} (1998) 127.
\bibitem{virpho1} J. Bartels, A. De Roeck, H. Lotter, Phys. Lett. \textbf{B} \textbf{389} (1996) 742.
\bibitem{virpho2} S. J. Brodsky, F. Hautmann, and D. E. Soper, Phys. Rev. \textbf{D56} (1997) 6957;
Phys. Rev. Lett. \textbf{78} (1997) 803 .
\bibitem{cepc} CEPC design performance considerations, arXiv:1501.06854.
\bibitem{ilc} ILC, Basic Conceptual Design Report, http://www.linearcollider.org.
\bibitem{eic} EIC, Basic Conceptual Design Report, arXiv:1212.1701.
\bibitem{eicc} D. P. Anderle $et~al$. Frontiers of Physics \textbf{64701} (2021) 16 Issue (6).
\bibitem{lep1} J. Bartels, C. Ewerz, and R. Staritzbichler, Phys. Lett. \textbf{B} \textbf{492} (2000) 56.
\bibitem{lep2} A. Donnachie, S. S$\ddot{\mathrm{o}}$ldner-Rembold, J. Phys. \textbf{G} \textbf{26} (2000) 689.
\bibitem{lep3} S. J. Brodsky, V. S. Fadin, V. T. Kim, L. N. Lipatov, and G. B.
Pivovarov, JETP Lett. \textbf{70} (1999) 155.
\bibitem{nlb2} M. Ciafaloni, G. Camici, Phys. Lett. \textbf{B} \textbf{430} (1998) 349.
\bibitem{im1} S. Gieseke, Nucl. Phys. \textbf{B} \textbf{121} (2003) 42.
\bibitem{im2} J. Bartels, D. Colferai, S. Gieseke and A. Kyrieleis, Phys. Rev.
\textbf{D66} (2002) 094017.
\bibitem{im3} J. Bartels, Nucl. Phys. \textbf{B} \textbf{116} (2003) 126.
\bibitem{qiao} J. Bartels, S. Gieseke and C. F. Qiao, Phys. Rev.
\textbf{D63} (2001) 056014 [Erratum-ibid. \textbf{D65} (2002)
079902].
\bibitem{feynarts} T.Hahn, Comput. Phys. Commun.
\textbf{140} (2001) 418 .
\bibitem{feyncalc} R.Metig, M.B$\mathrm{\ddot{o}}$hm and A.Denner, Comput.
Phys. Commun. \textbf{64} (1991) 345 .
\bibitem{looptools}T.Haln and M.P$\mathrm{\acute{e}}$rez-Victoria, Comput.
Phys. Commun. \textbf{118} (1999) 153
\bibitem{denner}A.Denner and S.Dittmaier, Nucl.
Phys. \textbf{B} \textbf{658} (2003) 175 .
\bibitem{ellis}R.Keith Ellis, Giulia Zanderighi, JHEP
\textbf{0802} (2008) 002.


\end{thebibliography}
\end{document}